\documentclass[%
 reprint,
bibnotes,
 amsmath,amssymb,
prb,
longbibliography
]{revtex4-2} 
\usepackage{graphicx}
\usepackage{dcolumn}
\usepackage{bm}

\usepackage{hyperref}
\usepackage{xcolor}

\begin{document}

\preprint{APS/123-QED}

\title{Trapped mode control in metasurfaces composed of particles with the form birefringence property}

\author{Anton S. Kupriianov$^{1}$}
\author{Volodymyr I. Fesenko$^{2}$}
\author{Andrey B. Evlyukhin$^{3}$}
\email{a.b.evlyukhin@daad-alumni.de}
\author{Wei Han$^{1}$}
\author{Vladimir~R.~Tuz$^{4,5}$}
\email{tvr@jlu.edu.cn}
\affiliation{$^1$College of Physics, International Center of Future Science, Jilin University, 2699 Qianjin St., Changchun 130012, China}
\affiliation{$^2$Institute of Radio Astronomy of National Academy of Sciences of Ukraine, 4, Mystetstv St., Kharkiv 61002, Ukraine}
\affiliation{$^3$Institute of Quantum Optics, Leibniz Universit\"at Hannover, 30167 Hannover, Germany}
\affiliation{$^4$State Key Laboratory of Integrated Optoelectronics, College of Electronic Science and Engineering, International Center of Future Science, Jilin University, 2699 Qianjin Street, Changchun, 130012, China}
\affiliation{$^5$School of Radiophysics, Biomedical Electronics and Computer Systems, V.~N.~Karazin Kharkiv National University, 4, Svobody Square, Kharkiv 61022, Ukraine}

\date{\today}

\begin{abstract}
Progress in developing advanced photonic devices relies on introducing new materials, discovered physical principles, and optimal designs when constructing their components. Optical systems operating on the principles of excitation of extremely high-quality factor trapped modes (also known as the bound states in the continuum, BICs) are of great interest since they allow the implementation of laser and sensor devices with outstanding characteristics. In this paper, we discuss how one can utilize the anisotropic properties of novel materials (transition metal dichalcogenides, TMDs), particularly, the bulk molybdenum disulfide (MoS$_2$), to realize the excitation of trapped modes in dielectric metasurfaces. The bulk MoS$_2$ is a thin-film structure in which the light wave behaves the same way as that in the uniaxial anisotropic material with the form birefringence property. Our metasurface is composed of an array of disk-shaped nanoparticles (resonators) made of the MoS$_2$ material under the assumption that the anisotropy axis of MoS$_2$ can be tilted to the rotation axis of the disks. We perform a detailed analysis of eigenwaves and scattering properties of such anisotropic resonators as well as the spectral features of the metasurface revealing dependence of the excitation conditions of the trapped mode on the anisotropy axis orientation of the MoS$_2$ material used.
\end{abstract}

\maketitle

\section{Introduction}

Thanks to astonishing advances in technology, electrically thin layers of materials can be fabricated with subwavelength periodic patterns to create planar metamaterials (also known as metasurfaces). Such two-dimensional structures make it possible to control the phase, amplitude, frequency, angular momentum, and polarization state of electromagnetic waves in a predetermined way \cite{Chen_2016, Li_Nanophotonics_2018}. Due to abrupt phase changes on device surfaces rather than phases of light propagating inside the systems, this type of metamaterials can be ultra-thin and flat, which is highly desirable in integrated optics and photonics. The main task in this field is to design metasurfaces that provide the most efficient interaction of light with nanostructured matter. With the use of novel materials, like transition metal dichalcogenides (TMDs) \cite{Wang_RevModPhys_2018, Verre_NatNano_2019, Munkhbat_lpor_2022}, one can also expect to make such metasurfaces manageable \cite{Xiao_2020}.

Atomically thin TMDs of the form MX$_2$ (M = Mo, W; X = S, Se, Te) are a new class of semiconductor materials proposed for use in both fundamental physics exploration in two-dimensional systems and device applications. Among the TMDs, MoS$_2$ is an indirect bandgap semiconductor with large bandgap energy which can be transformed into a direct semiconductor in the monolayer limit \cite{Yan_ApplOpt_2015, Ermolaev_npj_2020}. A bulk MoS$_2$ is considered as a layer-by-layer stack of MoS$_2$ monolayers (thin films), where strong interactions only exist within the basal plane of covalent bonds. Therefore, the bulk MoS$_2$ is an anisotropic crystal, which is macroscopically characterized with a tensor-valued permittivity possessing in-plane $\varepsilon_\parallel$ and out-of-plane  $\varepsilon_\bot$ terms \cite{Islam_adpr_2021, Ermolaev_Nat_2021}, resembling the form birefringence property \cite{born_book,yariv_book}. In particular, these extraordinary properties of MoS$_2$ have been utilized in constructing anisotropic resonators for lasing systems \cite{Stockman_JApplPhys_2016}, plasmon-assisted \cite{Ren_OptExpress_2021, Wang_ChinOptLett_2022} and dielectric metasurfaces \cite{Srivastava_AdvOptMater_2017, Bucher_ACSPhotonics_2019, Muhammad_NanoLett_2021, Li_PhysRevB_2022, Zhang_OptLett_2022}. 

The optical features of dielectric metasurfaces depend on the excitation conditions of corresponding Mie-type modes \cite{evlyukhin2010optical, Kuznetsov_Science_2016} arising in resonators. In particular, in the spectra of dielectric metasurfaces, there are two basic resonant states which correspond to the excitation of the magnetic dipole and electric dipole modes whose position on the frequency scale depends on the aspect ratio of the particles \cite{evlyukhin2011multipole}. Moreover, when metasurfaces containing anisotropic particles are considered, an additional mode splitting may occur \cite{Khromova_Laser_2016}. Performing tuning mode conditions by changing the aspect ratio of dielectric nanoparticles allows one to realize flexible manipulation by resonant states \cite{evlyukhin2011multipole, Razmjooei_OptExpress_2021} of the nanostructure, and implement, for example, Huygens metasurfaces \cite{Decker_AdvOptMat_2015} where destructive interference between the electric and magnetic dipole modes of comparable strength suppresses backward scattering making such metasurfaces to be fully transparent.

Among different designs of dielectric metasurfaces, we further distinguish a particular class of structures that allow obtaining the strongest resonant response due to the excitation of so-called trapped modes \cite{fedotov2007sharp} (recently such modes are referred to as bound states in the continuum, BICs \cite{Koshelev_PhysRevLett_2018}). In metasurfaces, such modes are related to non-radiating (dark) eigenwaves which cannot couple with the field of incident radiation (continuum). The origin and feature of a particular trapped mode depend on the symmetry of a system and its spatial, geometric, and material parameters. In particular, a trapped mode can be excited in metasurfaces if their constitutive particles (resonators) are in some way perturbed \cite{Khardikov_2010}. In this case, a perturbation transforms inherently dark mode to a weakly radiative one when the spatial symmetry of the translation unit cell of the metasurface is broken. The cluster-based unit cell designs \cite{Kupriianov_PhysRevApplied_2019, Xiao_PhysRevB_2022} as well as the use of specific irradiation conditions (oblique incidence \cite{He_PhysRevB_2018}, near-field sources \cite{van2021unveiling}, etc.) are other possible ways to realize the excitation of trapped modes in metasurfaces.

From the point of view of electromagnetic theory, breaking the spatial symmetry of resonators forming a metasurface leads to the appearance of artificial anisotropy of their averaged constitutive parameters resulting in their anisotropic or bianisotropic electromagnetic response \cite{Evlyukhin_PhysRevB_2020}. The symmetry breaking enables the trapped mode coupling with the continuum transforming the dark mode into a quasi-trapped one. Despite the leakage of energy from the quasi-trapped mode due to such artificial anisotropy, the concentration of electromagnetic energy in the resonators can be maintained at a very high level, if the introduced perturbation is small \cite{Koshelev_PhysRevLett_2018, Kafesaki_apl_2021}. Nevertheless, technologically, producing nanoscale resonators with the required small perturbation is a rather complicated task. Evidently, this issue can be solved with the use of anisotropic substances, in particular, the MoS$_2$ material.

Although, generally, the mechanism of excitation of trapped modes in dielectric metasurfaces is well described \cite{Tuz_OptExpress_2018, Tuz_JApplPhys_2019, Evlyukhin_PhysRevB_2020}, the effect of anisotropy in nanoparticles on such states, as far as we know, has not yet been thoroughly studied. Therefore, in this paper, we combine the anisotropic properties of  bulk MoS$_2$ considered for particle fabrication and the geometric parameters of these particles to construct a metasurface and study the influence of the birefringence property on the resonant conditions of the trapped mode excitation. In particular, in our metasurface, we utilize disk-shaped resonators due to their amenability to modern lithography techniques used for producing metasurfaces operating in the near-infrared range. All our numerical simulations are targeted on the third telecommunication window \cite{Bi_AdvOptMater_2021}. In contrast to the results published earlier for the BIC-assisted metasurfaces composed of resonators made of TMDs \cite{evlyukhin2021polarization, Prokhorov_PhysRevB_2022, Prokhorov_acsphotonics_2022, qin2022strong}, here we consider resonators without any perturbation of their geometric symmetry when the metasurface is under the normal irradiation conditions. Our approach is based on the analysis of eigenwaves of such resonators and their electric and magnetic dipole responses revealing the dependence of the excitation conditions of the trapped mode on the anisotropy axis orientation in the bulk MoS$_2$ material.

\section{\label{sec:eigenwaves}Eigenwaves analysis}

\begin{figure}[t!]
\centering
\includegraphics[width=0.5\textwidth]{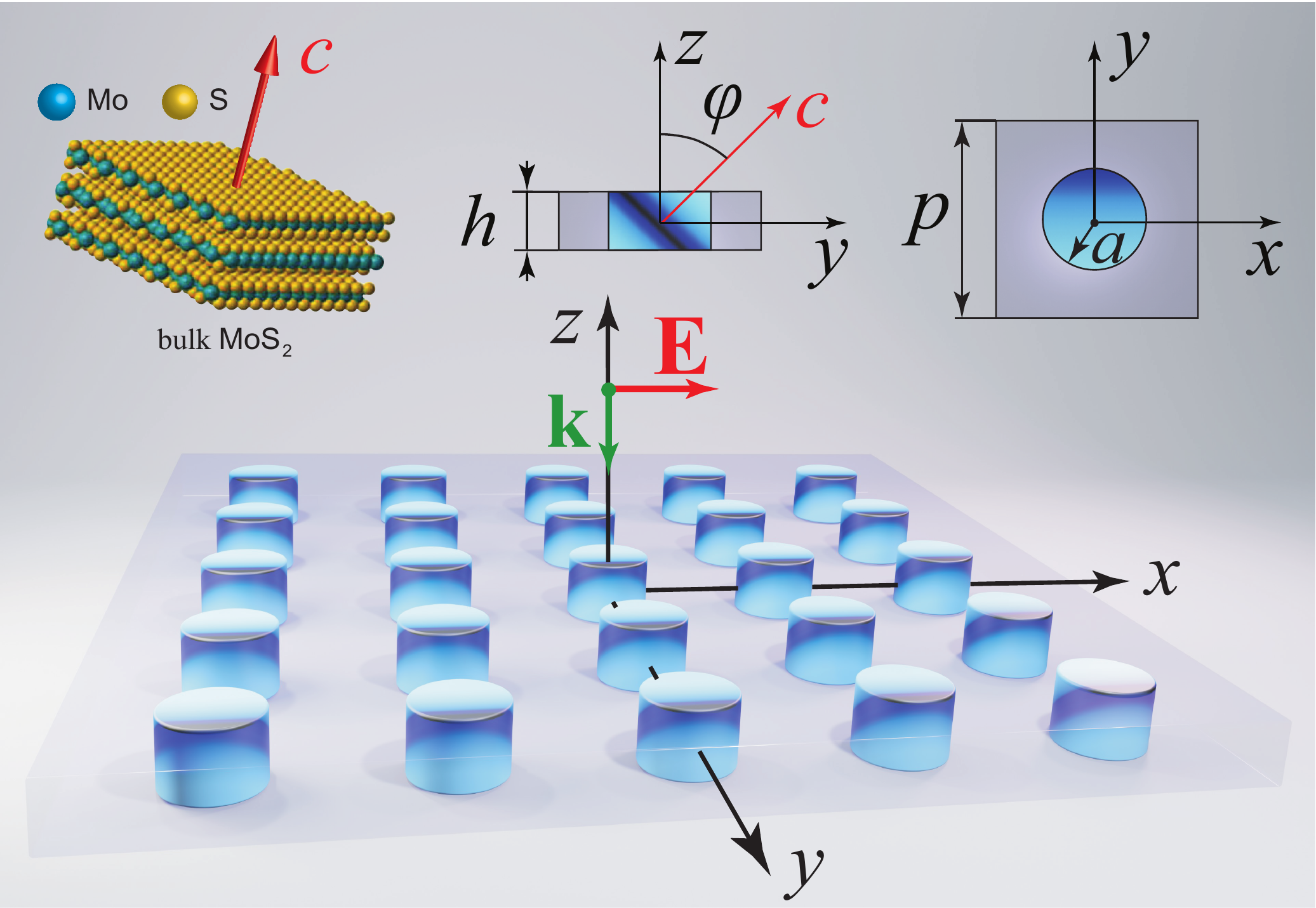}
\caption{Schematic illustration of the bulk MoS$_2$ material (thin-film structure) and sketch view of the metasurface made of MoS$_2$ anisotropic disk-shaped nanoparticles, the incident wave conditions, and geometric parameters of the metasurface unit cell. The $c$-axis tilt angle $\varphi$ of the MoS$_2$ anisotropic material is measured from the $z$-axis.}
\label{fig:fig1}
\end{figure}

In what follows, we consider a metasurface composed of MoS$_2$ anisotropic disk-shaped nanoparticles arranged in a two-dimensional array. The translation unit cell of this array is a square with the side size $p$. The radius and height of the disks are $a$ and $h$, respectively. The bulk MoS$_2$ material is a thin-film structure, which is characterized by tensor-valued permittivity $\hat\epsilon$. The light wave in such a material behaves the same way as that in the uniaxial anisotropic crystal with the form birefringence property \cite{born_book, yariv_book}. In particular, it is a nonmagnetic ($\mu$ = $\mu_0$) uniaxial anisotropic crystal whose anisotropy axis ($c$-axis) is directed perpendicular to the plane of layers forming the MoS$_2$ thin-film structure \cite{Islam_adpr_2021, Ushkov_PhysRevB_2022}. For the sake of definiteness, we assume that the $c$-axis of the crystal is tilted in the $y$-$z$ plane and can change its orientation on the angle $\varphi$ with respect to the $z$-axis of the chosen Cartesian coordinate frame, where the $z$-axis is also the rotation (axial symmetry) axis of the disks, as shown in Fig.~\ref{fig:fig1}. In the operating frequency range under study, material losses in the bulk MoS$_2$ material are negligibly small (see Appendix~\ref{appA}) and, therefore, we exclude them from our consideration. Without loss of generality, we assume that the particles are located in a homogeneous surrounding space (air). The trapped mode manifestation is related to the in-plane symmetry breaking of the metasurface unit cell. The substrate presence does not break the in-plane symmetry and has no significant influence on the effects discussed. Therefore, in order not to overload our study, we consider that the substrate is made of an air-like material $\varepsilon_s\approx 1$. Accounting for the substrate can be performed later in the engineering implementation of the metasurface by adjusting the geometric parameters of the resonators used.

\begin{figure*}[t]
\centering\includegraphics[width=1.0\textwidth]{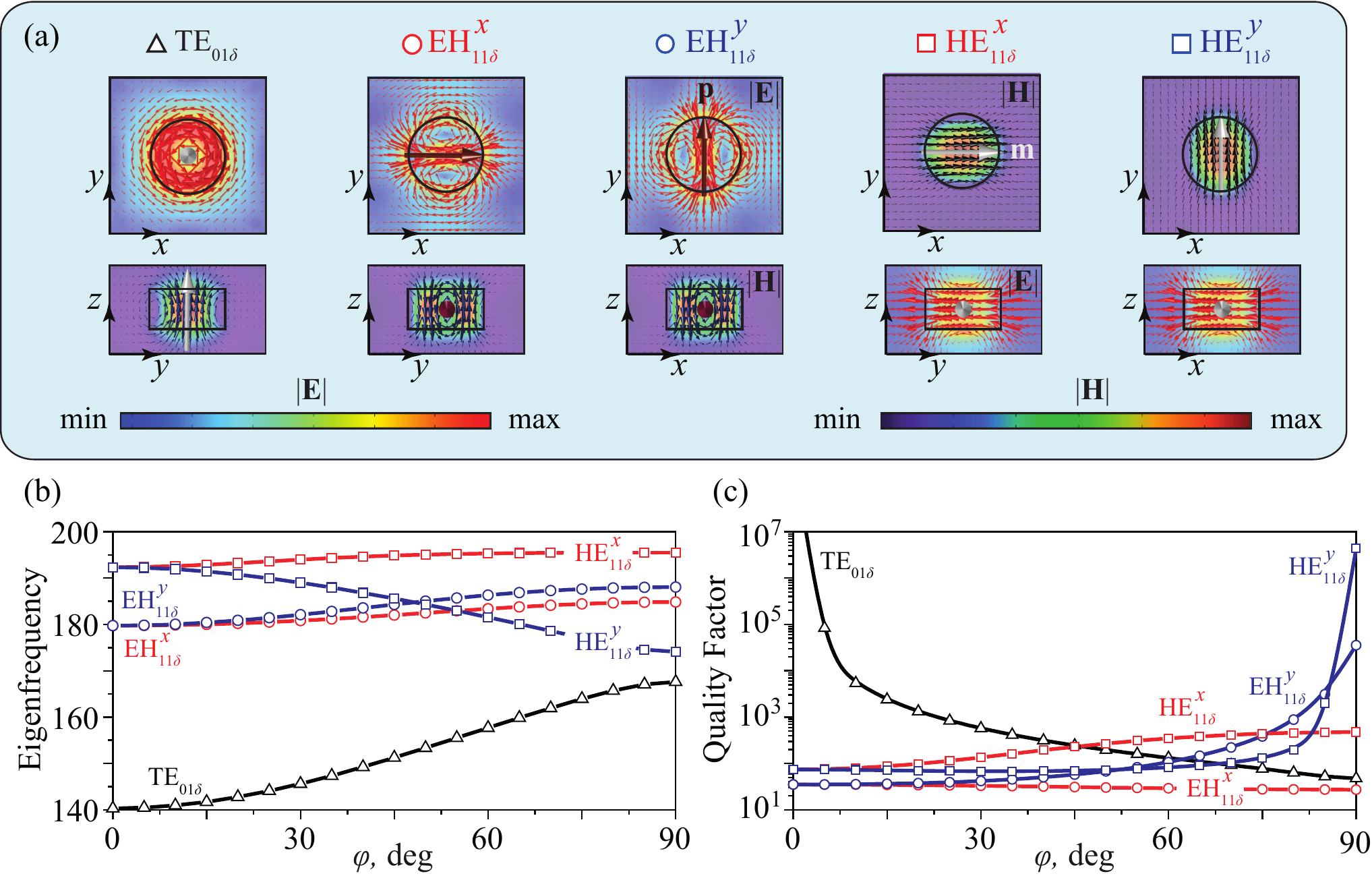} \caption{(a) Patterns of the electric and magnetic fields plotted for the $\textsc{TE}_{01\delta}$, $\textsc{EH}_{11\delta}^{x,y}$, and $\textsc{HE}_{11\delta}^{x,y}$ eigenwaves of the MoS$_2$  metasurface at $\varphi=0^\circ$, where the red and dark blue arrows represent the flow of electric and magnetic fields while brown and gray arrows represent moments of the electric dipole $\bf p$ and magnetic dipole $\bf m$, respectively. (b)~Eigenfrequencies ($\omega^\prime/2\pi$) and (c) quality factors ($\omega^\prime/2\omega^{\prime\prime}$) of eigenwaves as functions of the $c$-axis tilt angle $\varphi$. Parameters of the metasurface are: $a = 258$ nm, $h = 275$ nm, and $p$ = 1500 nm.}
\label{fig:fig2}
\end{figure*}

Initially, we perform an eigenwave analysis of the given metasurface. For our simulations, we use the RF module of the commercial COMSOL Multiphysics finite-element electromagnetic solver. In the solver, we construct one square unit cell of the metasurface imposing the Floquet-periodic boundary conditions on its sides to simulate the infinitely expanded two-dimensional array of nanoparticles. From the solutions found, we select those that correspond to the excitation of the lowest-order eigenwaves only. We classify these eigenwaves by considering the electromagnetic field configuration inside the resonator when the $c$-axis tilt angle $\varphi$ is zero and permittivity of the disk acquires the diagonal form $\hat\epsilon|_{\varphi=0^\circ}= \{ \varepsilon_\parallel, \varepsilon_\parallel, \varepsilon_\bot \}$. For the chosen geometric parameters of the resonators, the eigenwaves arise as the transverse electric TE$_{01\delta}$, and hybrid EH$_{11\delta}$ and HE$_{11\delta}$ modes, sequentially arranged with increasing frequency (this classification is based on the mode nomenclature of cylindrical dielectric waveguides, see Ref.~\cite{Snitzer_JOSA_1961}). In the subscripts of the mode abbreviations, the first index denotes the azimuthal variation of the fields, the second index denotes the order of variation of the field along the radial direction, and the third index denotes the order of variation of fields along the $z$-direction. Since we do not fix the exact number of field variations along the $z$-axis, we substitute the third index with the letter $\delta$. In general, the characteristics of eigenwaves under study inherit the properties of the corresponding modes of an anisotropic nanowire derived in Appendix~\ref{appB}. 

Moreover, as known from the theory of dielectric resonators \cite{Mongia_1994}, the TE$_{01\delta}$ mode of an isolated cylindrical resonator radiates like a magnetic dipole $\bf m$ oriented along its rotation axis (out-of-plane), whereas the EH$_{11\delta}$ and HE$_{11\delta}$ modes radiate like an electric dipole $\bf p$ and magnetic dipole $\bf m$, respectively, oriented along the transverse direction (in-plane). Thus, each hybrid mode is complemented by its own degenerated state, depending on the direction in which the corresponding dipole moment is oriented, along either the $x$-axis or the $y$-axis in the chosen Cartesian coordinate frame. We distinguish these degenerated hybrid modes by adding the corresponding superscript $x$ or $y$ to the mode abbreviations. The appearance of all eigenwaves of our interest at the initial stage $\varphi=0$ are collected in Fig.~\ref{fig:fig2}(a).

Among these eigenwaves, the field for the TE$_{01\delta}$ mode is axisymmetric and thus has no azimuthal variation, whereas the fields of the hybrid modes EH$_{11\delta}^{x,y}$ and HE$_{11\delta}^{x,y}$ are azimuthally dependent. This property of the TE$_{01\delta}$ mode leads to the appearance of conditions corresponding to the existence of the trapped mode in our lossless metasurface signifying that the corresponding eigenfrequency ($\omega = \omega^\prime + {\rm i}\omega^{\prime\prime}$) is a purely real quantity ($\omega^{\prime\prime}=0$).

Further in our study, we vary the value of the tilt angle $\varphi$. When $\varphi\in (0^\circ, 90^\circ)$, there are off-diagonal components in the tensor $\hat\epsilon$, whereas when $\varphi=90^\circ$, the $c$-axis is directed along the $y$-axis, and permittivity of the disk becomes diagonal $\hat\epsilon|_{\varphi=90^\circ}= \{ \varepsilon_\parallel, \varepsilon_\bot, \varepsilon_\parallel \}$ (see Appendix~\ref{appB}). We track the change in the values of the real part of the eigenfrequency ($\omega^\prime$) and quality factor ($\omega^\prime/ 2\omega^{\prime\prime}$) of the selected eigenwaves. The corresponding results of our calculations are presented in Figs. \ref{fig:fig2}(b) and \ref{fig:fig2}(c), respectively. One can see that the hybrid EH$_{11\delta}^{x,y}$ and HE$_{11\delta}^{x,y}$ modes lose their degeneracy and split apart with a change in the tilt angle $\varphi$ of the anisotropy axis, while the TE$_{01\delta}$ mode does not. Moreover, the quality factor of the TE$_{01\delta}$ mode changes from infinity to finite values undergoing a transformation from the trapped to a radiative state. Hereinafter, our task is to reveal the manifestation of these eigenwaves in the transmitted spectra of our metasurface.

\section{\label{sec:spectra}Spectral features}

\begin{figure*}[t]
\centering\includegraphics[width=1.0\textwidth]{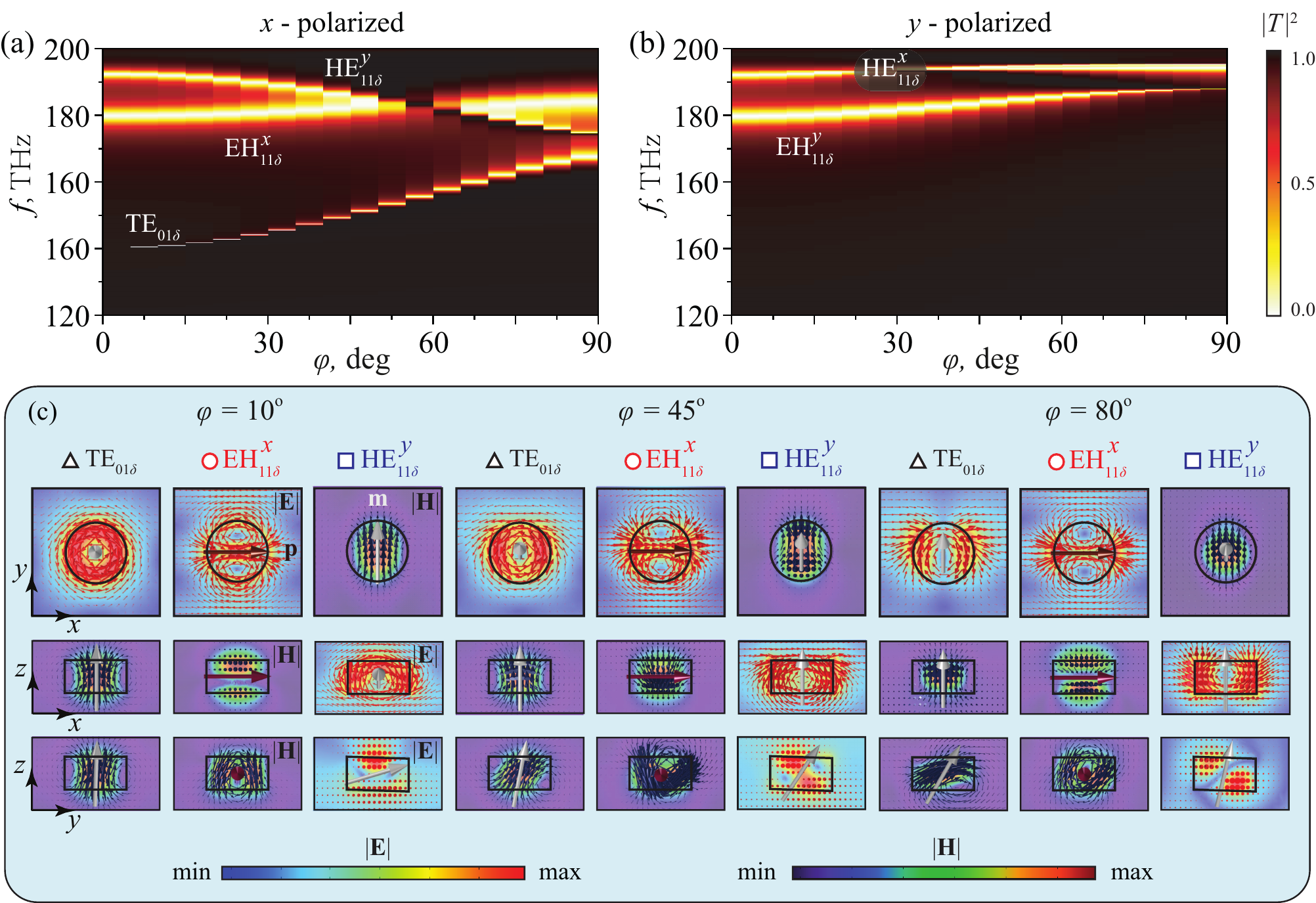} \caption{Transmission coefficient magnitude as a function of the frequency $f=\omega/2\pi$ and $c$-axis tilt angle $\varphi$ for the (a) $x$-polarized and (b) $y$-polarized incident waves. (c) Patterns of the electric and magnetic fields plotted at the corresponding resonant frequencies for the $\textsc{TE}_{01\delta}$, $\textsc{EH}_{11\delta}^{x}$, and $\textsc{HE}_{11\delta}^{y}$ modes of the metasurface at three different values of $\varphi$ and the $x$-polarized wave. The red and dark blue arrows represent the flow of electric and magnetic fields while the brown and gray arrows represent the electric $\bf p$ and magnetic $\bf m$ dipole moments, respectively. Parameters of the metasurface are the same as in Fig. \ref{fig:fig2}.}
\label{fig:fig3}
\end{figure*}

In this section, we reveal specific spectral features of the MoS$_2$ metasurface caused by the material anisotropy. For this study we consider that the metasurface is illuminated by a plane electromagnetic wave under the normal incidence conditions ($\textbf{k}=\{0, 0, -k_z\}$) with the electric field polarized along either the $x$-axis ($\textbf{E} = \{E_x,0,0\}$, $x$-polarization) or the $y$-axis ($\textbf{E} = \{0,E_y,0\}$, $y$-polarization). In the framework of the RF module of the COMSOL Multiphysics solver, we dispose of the radiating and receiving ports above and below the metasurface, respectively. The port boundary conditions are placed on the interior boundaries of the perfectly matched layers, adjacent to the air domains. Perfectly matched layers on the top and bottom of the unit cell absorb the excited wave from the radiating port and prevent unwanted re-reflection inside the computational domain. The port boundary conditions automatically determine the reflection and transmission characteristics of the metasurface in terms of S-parameters. After simulation, we retrieve values of the transmission coefficient ($|T|^2=|S_{21}|^2$) as a function of the frequency and $c$-axis tilt angle $\varphi$ for two orthogonal polarizations of the incident wave. The results of our calculations are summarized in Figs. \ref{fig:fig3}(a) and \ref{fig:fig3}(b) for the $x$-polarized and $y$-polarized waves, respectively. 

When $\varphi=0^\circ$, the metasurface is polarization-independent. In the frequency band of interest, there are two resonant states related to the excitation of the degenerated hybrid EH$_{11\delta}$ and HE$_{11\delta}$ modes. The identified resonances appear due to an existing electromagnetic coupling between the linearly polarized incident wave and transversely located electric and magnetic dipoles inherent to the corresponding eigenwaves. Since the TE$_{01\delta}$ mode is axially symmetric with the corresponding magnetic dipole oriented orthogonally to the metasurface plane, the field of the incident wave does not interact with this mode and there is no corresponding resonance in the spectra of the transmitted wave.

Variation of the $c$-axis tilt angle $\varphi$ breaks the in-plane symmetry of the structure, resulting that the metasurface becomes to be polarization-dependent. Although the degeneracy for hybrid modes is lifted, their resonant positions on the frequency scale change a little compared to those of the degenerated modes that existed in the $\varphi=0^\circ$ case. However, since the rotation of the anisotropy $c$-axis is maintained in the $y$-$z$ plane, noticeable changes are observed in the spectra of the $x$-polarized wave. At a certain value of the $c$-axis tilt angle ($\varphi\approx 50^\circ$), the resonant frequencies of the EH$_{11\delta}^x$ and HE$_{11\delta}^y$ modes coincide, and with a further increase in $\varphi$, the resonances rearrange their order on the frequency scale. This mode-crossing effect is widely discussed in the context of the implementation of Huygens metasurfaces \cite{Decker_AdvOptMat_2015}. Moreover, the quality factor of the EH$_{11\delta}^y$ and HE$_{11\delta}^y$ resonances increases, since the electromagnetic coupling of the corresponding either electric $\bf p$ or magnetic $\bf m$ dipole with the field of the incident wave decreases.

Aside from the resonances related to the above-considered hybrid EH$_{11\delta}$ and HE$_{11\delta}$ modes which possess a good electromagnetic coupling with the field of the normally incident linearly polarized wave, in the frequency band of interest, an additional resonance arises in the transmitted spectra of the $x$-polarized waves as soon as $\varphi$ is nonzero. This resonance is related to the manifestation of the TE$_{01\delta}$ mode. It appears as an alone-standing lowest frequency resonance whose quality factor decreases with increasing $\varphi$. This resonance acquires a peak-and-trough (Fano) profile as is typical for the trapped modes excitation \cite{fedotov2007sharp, Tuz_OptExpress_2018}. While the dip in curves corresponds to the maximum of reflection, the peak corresponds to the maximum of transmission. These extremes approach 0 and 1, respectively, since the material losses in the given metasurface are considered to be negligibly small.

To reveal the origin of changes in the emerging resonances, we plot the patterns of the electric and magnetic fields within the unit cell in three basic projections and show the orientation of the dipole moments $\bf m$ and $\bf p$ at the resonator center. These calculations are made for three different values of the $c$-axis tilt angle $\varphi$. The results of our calculations are collected in Fig. \ref{fig:fig3}(c). One can see that variation in the material anisotropy leads to a change in the slope of the dipole moment of the corresponding mode, thereby changing the degree of its electromagnetic coupling with the field of the incident wave. This mechanism is fully consistent with the behavior of trapped modes in resonators with perturbed unit cells \cite{Tuz_OptExpress_2018, Tuz_PhysRevB_2019, Pravin_AdvPhRes_2022} and the corresponding theory \cite{Tuz_JApplPhys_2019, Evlyukhin_PhysRevB_2020, Prokhorov_PhysRevB_2022} can be adapted in this case. It is based on the scattering characteristics of a single anisotropic nanoparticle which we discussed in Appendix~\ref{appC}.

\section{Polarization features}

\begin{figure*}[t]
\centering\includegraphics[width=1.0\textwidth]{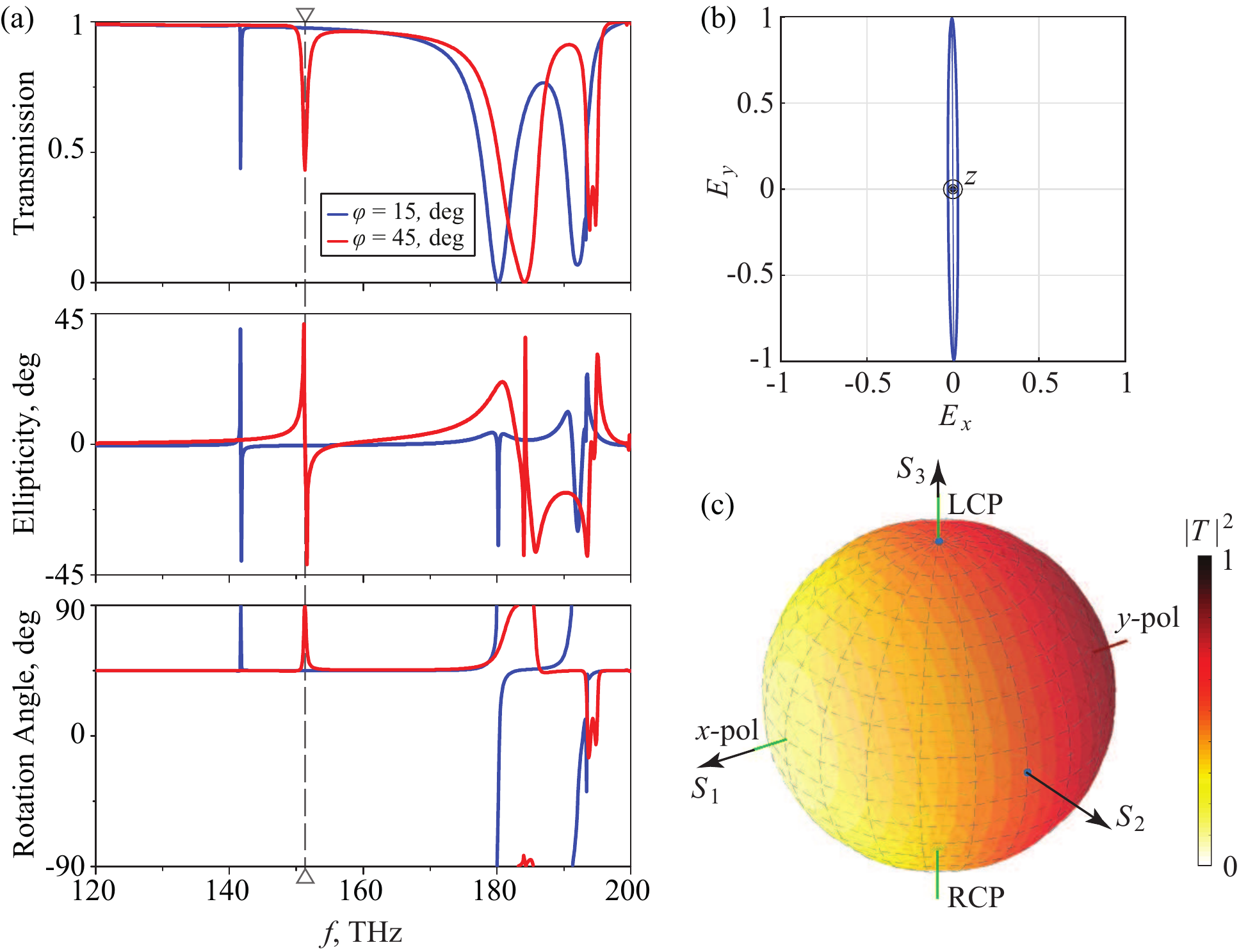} \caption{Polarization features of the transmitted field: (a) transmission coefficient magnitude of the circularly polarized waves, ellipticity, and rotation angle (polarization azimuth) as functions of the frequency $f=\omega/2\pi$ for two particular values of the $c$-axis tilt angle $\varphi$; (b) polarization ellipse and (c) transmission coefficient magnitude mapped on the surface of the Poincar\'e sphere related to the polarization states of the incident wave at the resonant frequency of the trapped mode excitation. The point at the north (south) pole of the Poincar\'e sphere represents the left-handed (right-handed) circularly polarized field, and the point on the equator represents the $45^\circ$ linearly polarized field. Parameters of the metasurface are the same as in Fig.~\ref{fig:fig2}.}
\label{fig:fig4}
\end{figure*}

Since the metasurface is composed of anisotropic particles, the polarization features of the metasurface at the resonant frequency of the trapped mode excitation should be elucidated. To this end, we study the transmission characteristics of our metasurface with respect to all polarization states of the incident wave for several values of the $c$-axis tilt angle $\varphi$. 

Figure~\ref{fig:fig4}(a) shows the corresponding frequency dependences of the transmission coefficient magnitude of circularly polarized waves, ellipticity angle ($\eta$), and polarization azimuth ($\theta$) for the transmitted field. According to the definition of the Stokes parameters, we introduce the ellipticity $\eta$ so that the field is linearly polarized when $\eta=0$, and $\eta=-\pi/4$ for the left-handed circularly polarized (LCP) wave and $+\pi/4$ for the right-handed circularly polarized (RCP) wave (note that in the latter cases the preferential azimuthal angle of the polarization ellipse $\theta$ becomes undefined). In all other cases ($0<|\eta|<\pi/4$), the field is elliptically polarized. In the considered frequency band, the transmission coefficient magnitude of the given metasurface is the same for the LCP and RCP incident waves, whereas the transmitted field undergoes the rotation of its polarization ellipse at the resonant frequencies where it can change its polarization state. However, at the resonant frequency of the trapped mode excitation, the transmitted field remains to be almost linearly polarized regardless of the value of the $c$-axis tilt angle $\varphi$. The corresponding polarization ellipse  of the transmitted field for the trapped mode resonance is presented in Fig.~\ref{fig:fig4}(b), whose frequency is depicted in Fig.~\ref{fig:fig4}(a) by a vertical dashed line.

This polarization feature of the trapped mode is further confirmed by our mapping of the transmission coefficient magnitude of the metasurface for different polarization states onto the surface of the full Poincar\'e sphere as shown in Fig.~\ref{fig:fig4}(c). In our calculations, the electric field vector ${\bf E}$ of the incident wave is defined by components $E_x=[0.5(P+1)]^{1/2}$ and $E_y=[0.5(P-1)]^{1/2}\exp(\rm i\beta)$, where $P\in [-1,1]$ and $\beta \in [-\pi, \pi]$ \cite{Tuz_PhysRevApplied_2020, Tuz_Nanophot_2021, Tuz_Advphot_2022}. The axes of the Poincar\'e sphere are labeled in terms of the Stokes parameters. Linearly polarized states are arranged along the sphere's equator, while right-handed (RCP) and left-handed (LCP) circularly polarized waves are located at their north and south poles, respectively. All other points of the sphere depict elliptically polarized states of the incident wave. From the uniform color gradient of the sphere, it is easy to conclude that at the resonant frequency of the trapped mode excitation, there are the total reflection and total transmission of $x$- and $y$-polarized incident waves, respectively, while circularly polarized waves pass through the metasurface with half the amplitude of the incident wave.

\section{Conclusions}

We have revealed the role of material anisotropy in the mechanism of the trapped mode excitation in a metasurface composed of MoS$_2$ disk-shaped nanoparticles. Our study is based on the analysis of the properties of the three lowest-order modes of cylindrical resonators forming the metasurface. They are the axially symmetric TE$_{01\delta}$ mode as well as the non-symmetric hybrid EH$_{11\delta}$ and HE$_{11\delta}$ modes. 

When the anisotropy axis of the MoS$_2$ material used for the disk-shaped particles fabrication coincides with the axis of their rotational symmetry, the hybrid EH$_{11\delta}$ and HE$_{11\delta}$ modes degenerate, whereas the TE$_{01\delta}$ mode behaves like a trapped mode in the metasurface, possessing a purely real eigenfrequency. When the anisotropy axis is tilted, the hybrid EH$_{11\delta}$ and HE$_{11\delta}$ modes are split, and the TE$_{01\delta}$ mode transforms into a radiative one, whose quality factor decreases with increasing the anisotropy tilt angle. It is shown how a change in the anisotropy tilt angle affects the orientation of vectors of the electric and magnetic dipole moments of the corresponding modes. 

While we have considered a specific MoS$_2$ material, the trapped mode excitation mechanism in metasurfaces composed of dielectric nanoparticles that we have demonstrated is general for both natural and artificial materials with uniaxial anisotropy. By performing control ability over the high-quality factor trapped modes in MoS$_2$ metasurfaces by manipulating material anisotropy, we propose a platform for many important applications in light-matter interaction, nonlinear photonics, quantum optics, and spectroscopy.

\acknowledgements{
V.R.T. is grateful for the hospitality and support from Jilin University, China. A.B.E. thanks funding support from the Deutsche Forschungsgemeinschaft (DFG, German Research Foundation) under Germany’s Excellence Strategy within the Cluster of Excellence PhoenixD (EXC 2122, Project ID 390833453).}

\appendix

\section{\label{appA}MoS$_2$ relative permittivity}

\begin{figure}[h]
\centering
\includegraphics[width=0.95\linewidth]{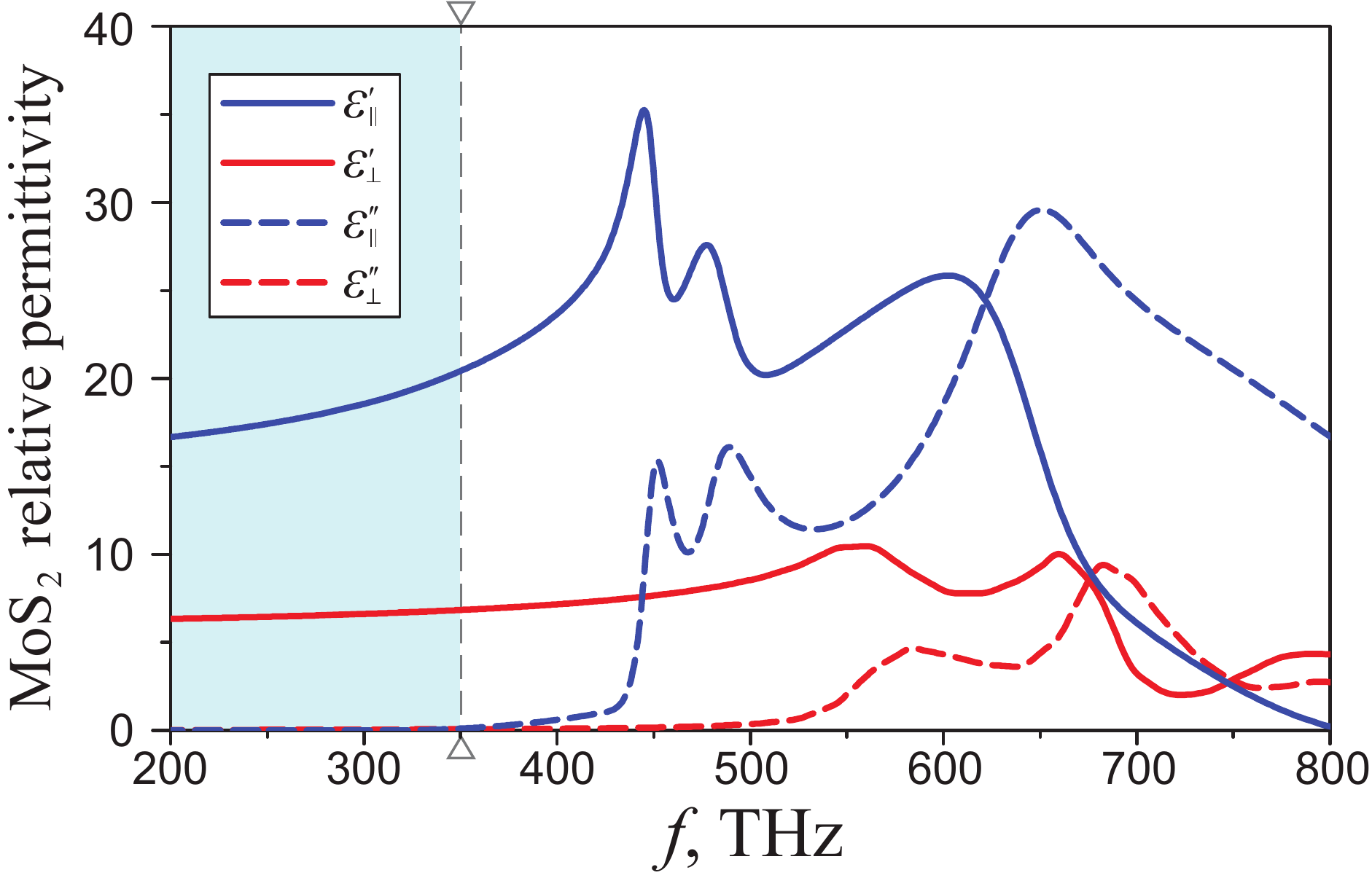}
\caption{Relative permittivity of MoS$_2$ extracted from experimental data of Ref.~\cite{Ermolaev_Nat_2021}.}
\label{fig:fig5}
\end{figure}

Dispersion characteristics of components of the tensor of the relative permittivity ($\hat\epsilon=\epsilon^\prime + {\rm i}\epsilon^{\prime\prime}$) of the MoS$_2$ material are shown in Fig.~\ref{fig:fig5}. Since in our work, we focus on the spectral range within the third telecommunication window ($\lambda \in [1500, 1600]~\rm{nm}$), where negligibly weak dispersion exists (see the spectral region marked light blue), we adopt $\varepsilon_{\parallel}^\prime=16.6$, $\varepsilon_{\perp}^\prime=6.3$, $\varepsilon_{\parallel}^{\prime\prime}=\varepsilon_{\perp}^{\prime\prime}=0.0$ in all considered region.

\section{\label{appB}Lowest-order modes of an anisotropic nanowire}

\begin{figure*}
\centering\includegraphics[width=1.0\textwidth]{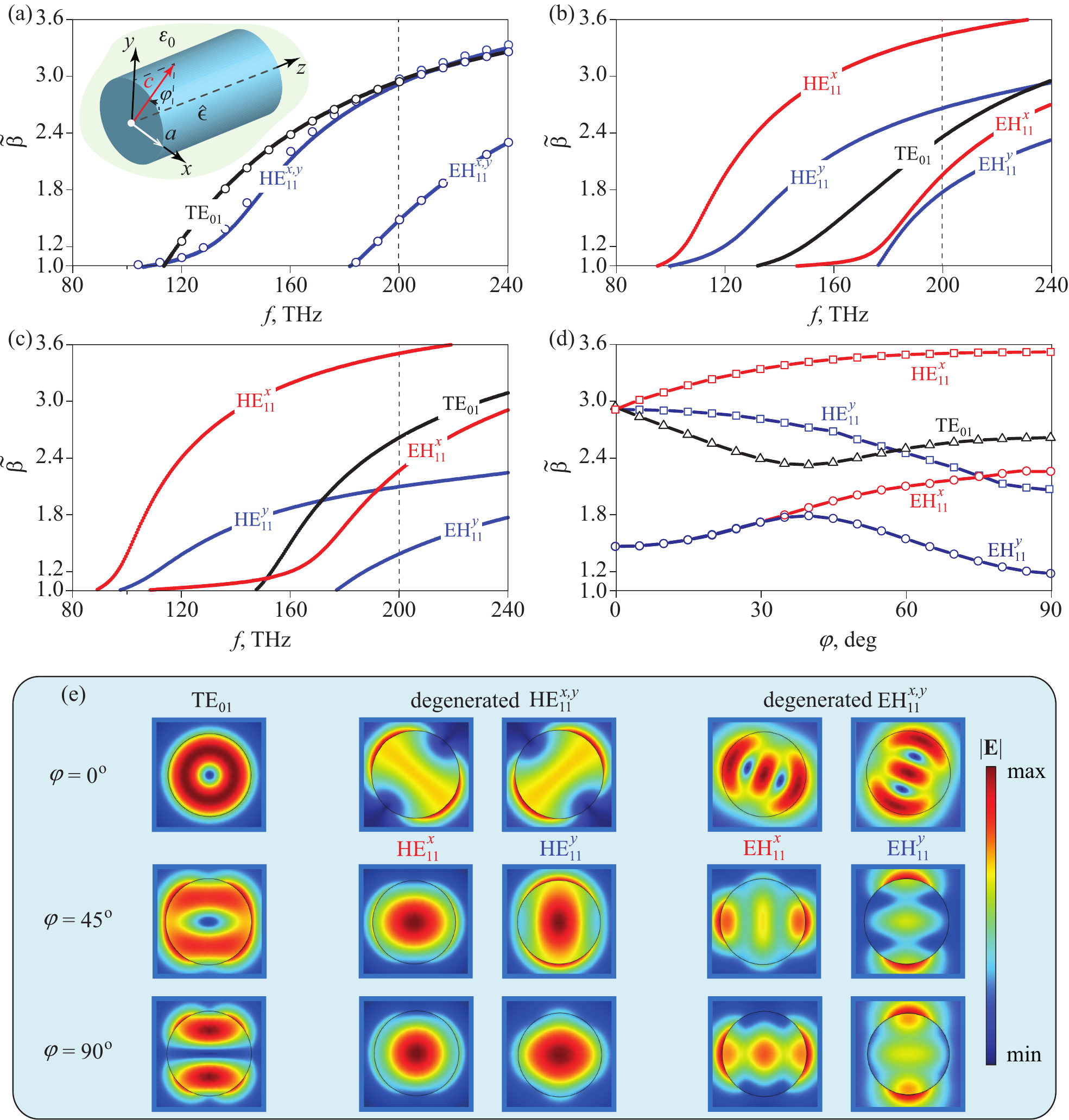} \caption{(a)-(c) Dispersion curves of the axially symmetric $\textsc{TE}_{01}$ and non-symmetric hybrid $\textsc{HE}_{11}^{x,y}$, $\textsc{EH}_{11}^{x,y}$ modes of the cylindrical MoS$_2$ nanowire at different values of the $c$-axis tilt angle $\varphi$: (a) $\varphi=0^\circ$, (b) $\varphi=45^\circ$, and (c) $\varphi=90^\circ$. The circles in panel (a) correspond to the analytical solution given by Eq.~(\ref{eq:DispEqNw}). (d) The effective mode index $\tilde{\beta}$ of the $\textsc{TE}_{01}$, $\textsc{HE}^{x,y}_{11}$ and $\textsc{EH}^{x,y}_{11}$ modes versus the $c$-axis tilt angle $\varphi$ at the fixed frequency $f=200.0$~THz. (e) Simulated electric field intensities $|\textbf{E}|$ of the nanowire modes corresponding to panels (a), (b), and (c). A schematic of the MoS$_2$ nanowire is shown in the inset of panel (a). The red arrow represents the $c$-axis direction in the nanowire core. The nanowire radius is $a=258$~nm, and components of the tensor of relative permittivity are: $\varepsilon_\parallel=16.6$ and  $\varepsilon_\bot=6.24$.}
\label{fig:fig6}
\end{figure*}

For reference, here we present dispersion features of three lowest-order modes (TE$_{01}$, EH$_{11}$, and HE$_{11}$) of an anisotropic nanowire. The nanowire is made of a nonmagnetic material (MoS$_2$) with the tensor-valued relative permittivity $\hat\epsilon$. The nanowire is disposed of in a surrounding medium with permittivity $\varepsilon_0$. The radius of the nanowire is $a$. The symmetry axis of the nanowire coincides with the $z$-axis of the Cartesian coordinate system [see the inset in Fig.~\ref{fig:fig6}(a)]. 

The optic $c$-axis of the uniaxial crystalline MoS$_2$ core of the nanowire is tilted in the $y$-$z$-plane and makes an angle $\varphi$ with the $z$-axis. The full relative permittivity tensor of the nanowire core $\hat\epsilon(\varphi)$ can be evaluated by using the rotation matrix 
\begin{equation}
\hat R =\left( {\begin{matrix}
   1 & 0 & 0 \cr
   0 & {\cos{\varphi}} & {-\sin{\varphi}} \cr
   0 & {\sin{\varphi}} & {\cos{\varphi}}\cr
\end{matrix}
} \right)
\label{eq:RM}
\end{equation}
as follows
\begin{widetext}
\begin{equation}
\hat\epsilon(\varphi) = \hat R \cdot \hat\epsilon|_{\varphi=0}\cdot \hat R^T=
\left( {\begin{matrix}
   \varepsilon_{xx} & 0 & 0 \cr
   0 &  {\varepsilon_{yy}\cos^2{\varphi}+\varepsilon_{zz}\sin^2{\varphi}} &  {\varepsilon_{yy}\sin{\varphi}\cos{\varphi}-\varepsilon_{zz}\sin{\varphi}\cos{\varphi}}\cr
   0 & {\varepsilon_{yy}\sin{\varphi}\cos{\varphi}-\varepsilon_{zz}\sin{\varphi}\cos{\varphi}} &  {\varepsilon_{zz}\cos^2{\varphi}+\varepsilon_{yy}\sin^2{\varphi}}\cr
\end{matrix}
} \right),
\label{eq:ep_r}
\end{equation}
\end{widetext}
where $\hat\epsilon|_{\varphi=0^\circ}= \{ \varepsilon_{xx}, \varepsilon_{yy}, \varepsilon_{zz} \}=\{ \varepsilon_\parallel, \varepsilon_\parallel, \varepsilon_\bot \}$ and superscript $T$ denotes the matrix transposition operation.

As follows from Eq.~(\ref{eq:ep_r}), the permittivity of the nanowire can be defined in the diagonal form as $\hat\epsilon|_{\varphi=0^\circ}= \{ \varepsilon_\parallel, \varepsilon_\parallel, \varepsilon_\bot \}$ and $\hat\epsilon|_{\varphi=90^\circ}= \{ \varepsilon_\parallel, \varepsilon_\bot, \varepsilon_\parallel \}$ in two special cases when the $c$-axis is directed along the $z$-axis and the $y$-axis, respectively. For the rest range  $\varphi\in\left(0^\circ,90^\circ\right)$, the tensor $\hat\epsilon(\varphi)$ has nonzero off-diagonal components.

When the directions of the $c$-axis and the $z$-axis coincide, there is an analytical solution for the nanowire modes \cite{Tonning_MTT_1982}. Generally, such an anisotropic nanowire supports an infinite set of the axially symmetric TE$_{mn}$ and TM$_{mn}$ modes as well as the non-symmetric hybrid EH$_{mn}$ and HE$_{mn}$ modes, whose dispersion equation can be expressed as \cite{Paul_RS_1981}:
\begin{equation}
\begin{split}
&\left[ \frac{\varepsilon_0}{\gamma a}\frac{K'_m(\gamma a)}{K_m(\gamma a)} + \frac{\sqrt{\varepsilon_\parallel \varepsilon_\bot}}{\kappa a} \frac{J'_m(p \kappa a)}{J_m(p \kappa a)} \right]\\
&\times\left[ \frac{1}{\gamma a}\frac{K'_m(\gamma a)}{K_m(\gamma a)} + \frac{1}{\kappa a} \frac{J'_m(\kappa a)}{J_m(\kappa a)} \right] \\
&=\left( \frac{m\beta}{k_0}\right)^2
\left[\frac{1}{(\gamma a)^2}+\frac{1}{(\kappa a)^2}\right]^2,
\end{split}
\label{eq:DispEqNw} 
\end{equation}
where, $k_0$ is the wavenumber in free space, $m$ and $n$ are the integer values known as the azimuthal and radial indices, respectively, $\beta$ is the longitudinal propagation constant, $p = \sqrt{\varepsilon_\bot / \varepsilon_\parallel}$ is the anisotropy parameter, $\kappa = \sqrt{k_0^2\varepsilon_\parallel-\beta^2}$ and $\gamma = \sqrt{\beta^2-k_0^2\varepsilon_0}$ are the transverse propagation constants inside and outside the nanowire, respectively, $J_m(\cdot)$ and $K_m(\cdot)$ are the Bessel function of the first kind and the modified Bessel function of the second kind, and $J'_m(\cdot)$ and $K'_m(\cdot)$ are their derivatives with respect to the function argument.

\begin{figure*}[t]
\centering\includegraphics[width=1.0\textwidth]{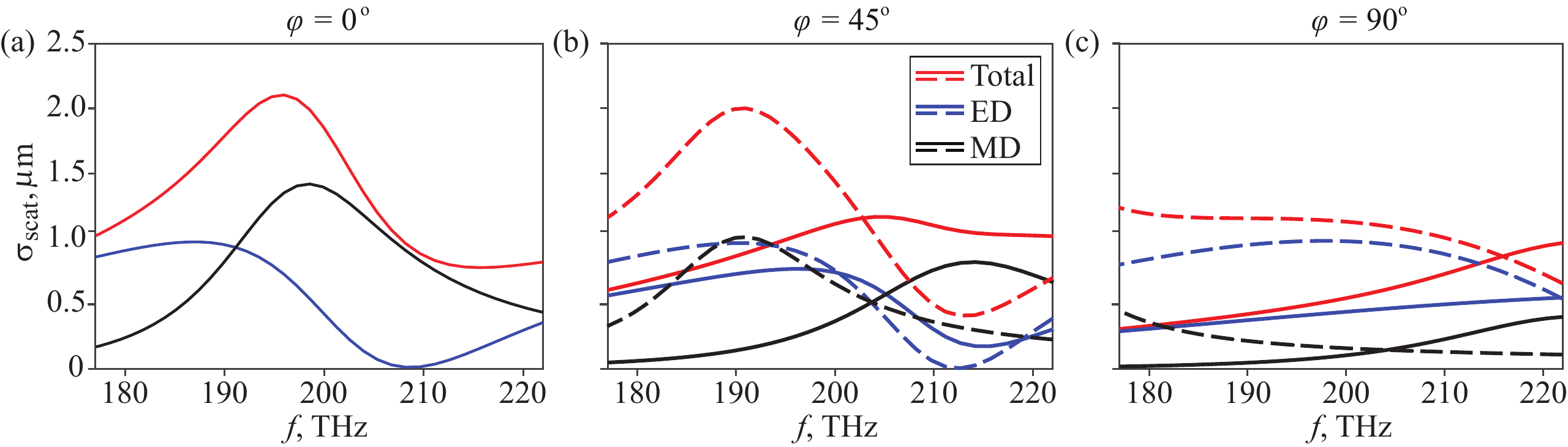} \caption{Contributions of the electric dipole (ED), magnetic dipole (MD), and total scattering cross-section of a single anisotropic MoS$_2$ nanoparticle (disk) for three different orientations of the $c$-axis of anisotropy with respect to the rotation axis of the disk, where (a) $\varphi=0^\circ$, (b) $\varphi=45^\circ$, and (c) $\varphi=90^\circ$, and the solid and dashed lines correspond to the frontal disk irradiation by the $x$-polarized and $y$-polarized wave, respectively.  Geometric parameters of the disk are: $a=258$~nm, $h=275$~nm, and components of the tensor of relative permittivity are: $\varepsilon_\parallel=16.6$ and  $\varepsilon_\bot=6.24$.}
\label{fig:fig7}
\end{figure*}

A formal substitution of $m=0$ into Eq.~(\ref{eq:DispEqNw}) yields us two separate dispersion equations related to the transverse magnetic TM$_{0n}$ and transverse electric TE$_{0n}$ modes [see, the first and second multiplier on the left side of the dispersion equation (\ref{eq:DispEqNw}), respectively], similarly to the case of a nanowire with an isotropic core \cite{Yu_Nanophot_2018}.

In the general case of a uniaxial crystal core when the $c$-axis does not coincide with the $z$-axis, the propagation constants and field components can be determined numerically \cite{Vandenbulcke_EL_1976, Koshiba_JLT_1986, Bagatskaya_TRE_2005}. The calculated effective mode index $\tilde{\beta}=\beta/k_0$ and the electric field patterns of the lowest-order $\textsc{TE}_{01}$, $\textsc{EH}^{x,y}_{11}$, and $\textsc{HE}_{11}^{x,y}$ modes propagated in the MoS$_2$ nanowire are presented in Fig.~\ref{fig:fig6} for different values of the $c$-axis tilt angle $\varphi$. In particular, when the $c$-axis is directed along the $z$-axis, the results of our simulation are checked against those obtained from the analytical solution given by Eq.~(\ref{eq:DispEqNw}).

From Figs.~\ref{fig:fig6}(a)-\ref{fig:fig6}(c), one can conclude that when $\varphi\in\left[0^\circ,90^\circ\right]$, the dispersion characteristic of the $\textsc{TE}_{01}$ mode does not depend on the transverse polarization, that is, this mode does not split into the $x$-polarized ($\textsc{TE}_{01}^x$) and $y$-polarized ($\textsc{TE}_{01}^y$) components (see, also Refs.~\cite{Dai_JOSA_1991, Dai_JOSA_1992}). Nevertheless, when $\varphi\neq0^\circ$, the axial symmetry of the electric field of the $\textsc{TE}_{01}$ mode becomes violated, as shown in Fig.~\ref{fig:fig6}(e).

In turn, the dispersion characteristics of hybrid modes are more complex. Figure~\ref{fig:fig6}(a) shows that at $\varphi=0^\circ$, no modal birefringence phenomenon exists, it means that the hybrid $\textsc{HE}^{x,y}_{11}$ and $\textsc{EH}^{x,y}_{11}$ modes are degenerated as shown in Fig.~\ref{fig:fig6}(e), which is typical for uniaxial nanowires \cite{Dai_JOSA_1991,Dai_JOSA_1992}.

In the range $\varphi\in(0^\circ,90^\circ]$, the dispersion characteristics of hybrid $\textsc{EH}^{x,y}_{11}$ and $\textsc{HE}^{x,y}_{11}$ modes depend on the transverse polarization orientations, which leads to degeneracy lifting as presented in Fig.~\ref{fig:fig6}(e). Thus, the different polarizations of $\textsc{HE}_{11}^x$ ($\textsc{EH}_{11}^x$) and $\textsc{HE}_{11}^y$ ($\textsc{EH}_{11}^y$) modes have different dispersion relations and propagate like different modes, as shown in Figs. \ref{fig:fig6}(b) and \ref{fig:fig6}(c). For the chosen parameters of anisotropy, the effect of the $c$-axis orientation is much stronger for the $y$-polarized modes compared to the $x$-polarized ones due to the fact that $\varepsilon_{yy}(\varphi)\in\left[\varepsilon_{\bot}, \varepsilon_{\parallel} \right]$ while $\varepsilon_{xx}(\varphi)= \varepsilon_{\parallel}$ [see, for clarity, Eq.~(\ref{eq:ep_r})]. A similar feature has been reported previously for an anisotropic rectangular waveguide \cite{Koshiba_JLT_1986}. Besides, Fig.~\ref{fig:fig6}(d) indicates that difference between the $x$-polarized and $y$-polarized modes increases as the $c$-axis tilt angle $\varphi$ rises, and the difference between $\textsc{EH}_{11}^x$ and $\textsc{EH}_{11}^y$ modes is smaller than the difference between the $\textsc{HE}_{11}^x$ and $\textsc{HE}_{11}^y$ modes. 

\section{\label{appC}Scattering cross-section of an~anisotropic nanoparticle}

\begin{figure*}[t]
\centering\includegraphics[width=0.9\textwidth]{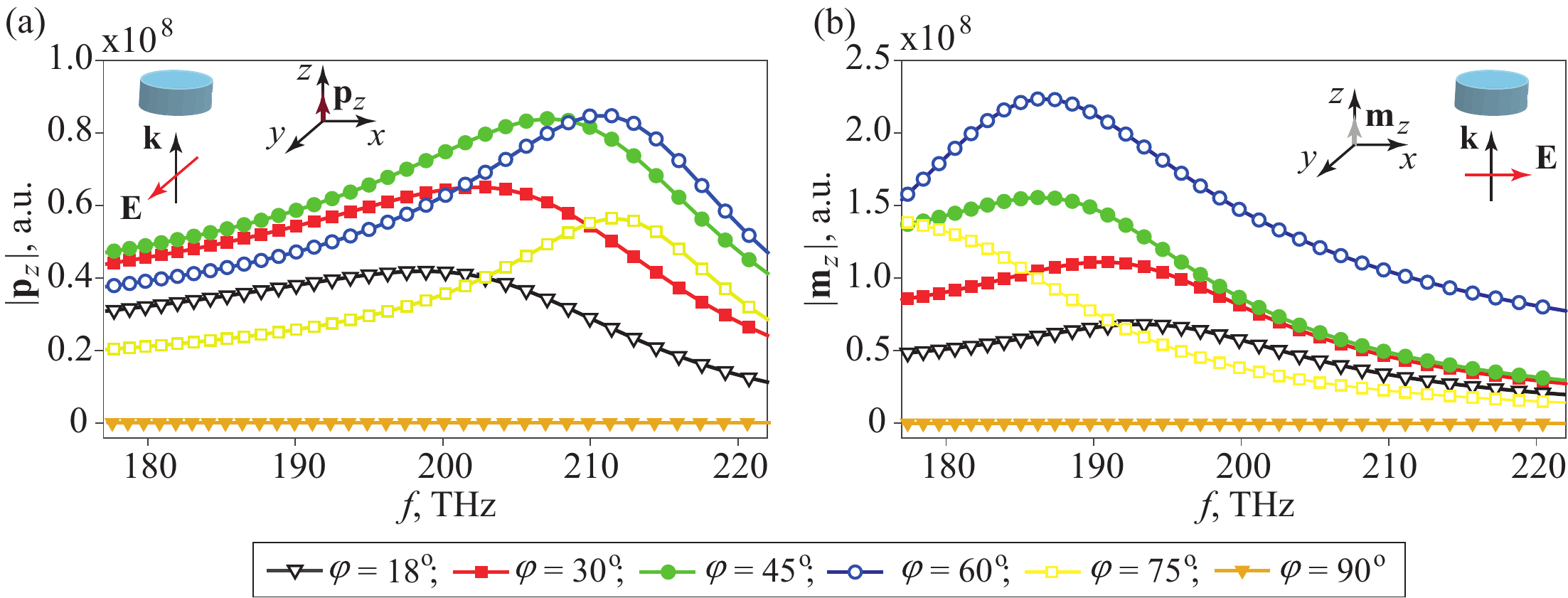} \caption{Variation of the magnitude of the $z$ component of the electric $\bf p$ and magnetic $\bf m$ dipole moments for several values of the $c$-axis tilt angle $\varphi$ for the frontal irradiation of the disk by the (a) $x$-polarized and (b) $y$-polarized wave. Parameters of the disk are the same as in Fig. \ref{fig:fig7}.}
\label{fig:fig8}
\end{figure*}

For numerical simulations of the single particle scattering cross-section and the particle electric and magnetic dipole moments, we use the discrete dipole approximation (DDA) and the decomposed discrete dipole approximation (DDDA) methods \cite{evlyukhin2011multipole}. The main idea of the DDA method consists in the replacement of the scattering object by a cubic lattice of electric point dipoles with known polarizability tensor $\hat\alpha_p$. The corresponding dipole moment ${\bf d}_i$ induced in each lattice point $i$ (with the radius vector ${\bf r}_i$) is found by solving coupled dipole equations:
\begin{equation}\label{ddd}
    {\bf d}^i=\hat\alpha_p{\bf E}^i_0+\hat\alpha_p\frac{k_0^2}{\varepsilon_0}\sum_{j\neq i}^N\hat G_{ij} {\bf d}^j,\quad i=1...N,
\end{equation}
where ${\bf E}^i_0$ is the electric field of the incident wave at the point ${\bf r}_i$, $k_0$ is the wavenumber in a vacuum, $\varepsilon_0$ is the vacuum dielectric constant, $\hat G_{ij}=\hat G({\bf r}_i,{\bf r}_j)$ is the Green tensor of the system without the particle, $N$ is the number of dipoles replacing the particle after the discretization procedure. For a MoS$_2$ particle (disk), when the $c$-axis of anisotropy is directed along the $z$-axis in the chosen Cartesian coordinate frame ($\varphi=0$), the permittivity tensor is written as 
\begin{equation}
\label{eq:epsp}
    \hat\epsilon=\left(
    \begin{array}{ccc}
         \varepsilon_{\parallel}&0&0  \\
        0 & \varepsilon_{\parallel}&0\\
        0 & 0& \varepsilon_{\bot}
    \end{array}
    \right),
\end{equation}
and the polarizability tensor can be taken as
\begin{equation}\label{al}
    \hat\alpha_p\simeq3\varepsilon_0\varepsilon_s V_p\left(
    \begin{array}{ccc}
         \frac{\varepsilon_{\parallel}-\varepsilon_s}{\varepsilon_{\parallel}+2\varepsilon_s}&0&0  \\
        0 & \frac{\varepsilon_{\parallel}-\varepsilon_s}{\varepsilon_{\parallel}+2\varepsilon_s}&0\\
        0 & 0& \frac{\varepsilon_{\bot}-\varepsilon_s}{\varepsilon_{\bot}+2\varepsilon_s}
    \end{array}
    \right),
\end{equation}
where $V_p$ is the volume of the discretization cell, $\varepsilon_s$ is the relative permittivity of the surrounding medium. Here we assume that the particle is localized in a vacuum or in the air where $\varepsilon_s=1$. After the solution of Eqs.~(\ref{ddd}) and choosing the origin of the Cartesian coordinate system in the center of mass of the particle,  the scattered electric field ${\bf E}^\textrm{scat}$ in the far-field zone and the spherical coordinates are written as  
\begin{eqnarray}
{E}^\textrm{scat}_\phi(r,\phi,\theta)&=&\frac{k_0^2e^{{\rm i}k_s r}}{4\pi\varepsilon_0 r}\sum_{j=1}^N e^{-{\rm i}k_s({\bf n}\cdot{\bf r}_j)}\left(d^j_y\cos\phi\right.\nonumber\\
    &&\left.-d_x^j\sin\phi\right),\nonumber\\
    {E}^\textrm{scat}_\theta (r,\phi,\theta)&=&\frac{k_0^2e^{{\rm i}k_s r}}{4\pi\varepsilon_0r}\sum_{j=1}^N e^{-{\rm i}k_s({\bf n}\cdot{\bf r}_j)}\left(d^j_x\cos\phi\cos\theta\right.\nonumber\\
    &&\left.+d_y^j\sin\phi\cos\theta-d_z^j\sin\theta\right),
\end{eqnarray}
where $k_s=k_0\sqrt{\varepsilon_s}$, ${\bf n}=(\sin\theta\cos\phi,~\sin\theta\sin\phi,~\cos\theta)$ is the unit vector directed to the observation point with spherical coordinate $(r,~\phi,~\theta)$, $r$ is the distance to the observation point, $\phi$ and $\theta$ are the azimuthal and polar angles, respectively.

The scattering cross-section $\sigma_\textrm{scat}$ is calculated as 
\begin{equation}
    \sigma_\textrm{scat}=\frac{1}{|{\bf E}_0|^2}\int_{\Omega}\left\{|{E}^\textrm{scat}_\phi|^2+|{E}^\textrm{scat}_\theta|^2\right\}r^2d\Omega,
\end{equation}
where $\Omega$ and $d\Omega=\sin\theta d\phi d\theta$ is the full solid angle and its differential.

In the DDDA, a discrete representation of the induced polarization inside the particle is obtained as
\begin{equation}\label{Pol}
    {\bf P}({\bf r})=\sum_{j=1}^N{\bf d}^j\delta({\bf r}-{\bf r}_j),
\end{equation}
where $\delta({\bf r}-{\bf r}_j)$ is the Dirac delta function.

Using the relation ${\bf j}=-{\rm i}\omega{\bf P}$, where $\omega$ is the field angular frequency, between the induced polarization $\bf P$ and the induced electric current density $\bf j$ and Eq. (\ref{Pol}), the multipole moments of the particle can be calculated by applying the expressions from Ref. \cite{evlyukhin2019multipole}. In this approach the multipole decomposition of the scattering cross-section is
\begin{eqnarray}
    \sigma_\textrm{scat}&&\simeq\frac{k_0^4}{6\pi\varepsilon_0^2|{\bf E}_0|^2}|{\bf p}|^2+\frac{k_0^4\varepsilon_s\mu_0}{6\pi\varepsilon_0|{\bf E}_0|^2}|{\bf m}|^2\nonumber\\
    &&+\frac{k_0^6\varepsilon_s}{720\pi\varepsilon_0^2|{\bf E}_0|^2}\sum_{\alpha\beta}|Q_{\alpha\beta}|^2 +\frac{k_0^6\varepsilon_s^2\mu_0}{80\pi\varepsilon_0|{\bf E}_0|^2}\sum_{\alpha\beta}|M_{\alpha\beta}|^2,\nonumber\\
\end{eqnarray}
where $\mu_0$ is vacuum permeability, $\bf p$ and $\bf m$ are the electric and magnetic dipole moments, $\hat Q$ and $\hat M$ are the electric and magnetic quadrupole moments of the particle.

Note that from the practical point of view, it is convenient to rotate the disk under the fixed polarizability tensor (\ref{al}) in the numerical simulation in order to change the anisotropic properties of the disk.

The results of our modeling of the cross-section characteristics of a single particle for several different values of the $c$-axis tilt angle $\varphi$ are shown in Fig.~\ref{fig:fig7}. First, one can see that the resonant response of the disk depends on both orientations of the MoS$_2$ layers and irradiation conditions. Second, in the chosen frequency range, the resonant features are associated only with the resonant excitation of either the electric (ED) or magnetic (MD) dipole moments of the particle. Importantly, in the range, $\varphi\in(0,90^{\circ})$, the material anisotropy of the disk leads to the resonant excitation of the longitudinal (directed along the wave vector $\bf k$ of the incident wave)  components of the ED and MD dipole moments (Fig.~\ref{fig:fig8}). The excitation of $p_z$ ($m_z$) component is realized for the $y$-polarized ($x$-polarized) waves, respectively, and spectral positions of their extremes are completely correlated with the resonant features of the transmission spectra of the MoS$_2$ metasurface presented in  Figs.~\ref{fig:fig3}(a) and \ref{fig:fig3}(b), respectively. This demonstrates that the discussed resonant states of the anisotropic metasurface are basically influenced by the anisotropy and modes of an individual dielectric resonator rather than by the electromagnetic coupling between them.

\bigskip

\bibliography{anisotropic}

\begin{thebibliography}{59}%
\makeatletter
\providecommand \@ifxundefined [1]{%
 \@ifx{#1\undefined}
}%
\providecommand \@ifnum [1]{%
 \ifnum #1\expandafter \@firstoftwo
 \else \expandafter \@secondoftwo
 \fi
}%
\providecommand \@ifx [1]{%
 \ifx #1\expandafter \@firstoftwo
 \else \expandafter \@secondoftwo
 \fi
}%
\providecommand \natexlab [1]{#1}%
\providecommand \enquote  [1]{``#1''}%
\providecommand \bibnamefont  [1]{#1}%
\providecommand \bibfnamefont [1]{#1}%
\providecommand \citenamefont [1]{#1}%
\providecommand \href@noop [0]{\@secondoftwo}%
\providecommand \href [0]{\begingroup \@sanitize@url \@href}%
\providecommand \@href[1]{\@@startlink{#1}\@@href}%
\providecommand \@@href[1]{\endgroup#1\@@endlink}%
\providecommand \@sanitize@url [0]{\catcode `\\12\catcode `\$12\catcode
  `\&12\catcode `\#12\catcode `\^12\catcode `\_12\catcode `\%12\relax}%
\providecommand \@@startlink[1]{}%
\providecommand \@@endlink[0]{}%
\providecommand \url  [0]{\begingroup\@sanitize@url \@url }%
\providecommand \@url [1]{\endgroup\@href {#1}{\urlprefix }}%
\providecommand \urlprefix  [0]{URL }%
\providecommand \Eprint [0]{\href }%
\providecommand \doibase [0]{https://doi.org/}%
\providecommand \selectlanguage [0]{\@gobble}%
\providecommand \bibinfo  [0]{\@secondoftwo}%
\providecommand \bibfield  [0]{\@secondoftwo}%
\providecommand \translation [1]{[#1]}%
\providecommand \BibitemOpen [0]{}%
\providecommand \bibitemStop [0]{}%
\providecommand \bibitemNoStop [0]{.\EOS\space}%
\providecommand \EOS [0]{\spacefactor3000\relax}%
\providecommand \BibitemShut  [1]{\csname bibitem#1\endcsname}%
\let\auto@bib@innerbib\@empty
\bibitem [{\citenamefont {Chen}\ \emph {et~al.}(2016)\citenamefont {Chen},
  \citenamefont {Taylor},\ and\ \citenamefont {Yu}}]{Chen_2016}%
  \BibitemOpen
  \bibfield  {author} {\bibinfo {author} {\bibfnamefont {H.-T.}\ \bibnamefont
  {Chen}}, \bibinfo {author} {\bibfnamefont {A.~J.}\ \bibnamefont {Taylor}},\
  and\ \bibinfo {author} {\bibfnamefont {N.}~\bibnamefont {Yu}},\ }\bibfield
  {title} {\bibinfo {title} {A review of metasurfaces: physics and
  applications},\ }\href {https://doi.org/10.1088/0034-4885/79/7/076401}
  {\bibfield  {journal} {\bibinfo  {journal} {Rep. Prog. Phys.}\ }\textbf
  {\bibinfo {volume} {79}},\ \bibinfo {pages} {076401} (\bibinfo {year}
  {2016})}\BibitemShut {NoStop}%
\bibitem [{\citenamefont {Li}\ \emph {et~al.}(2018)\citenamefont {Li},
  \citenamefont {Singh},\ and\ \citenamefont
  {Sievenpiper}}]{Li_Nanophotonics_2018}%
  \BibitemOpen
  \bibfield  {author} {\bibinfo {author} {\bibfnamefont {A.}~\bibnamefont
  {Li}}, \bibinfo {author} {\bibfnamefont {S.}~\bibnamefont {Singh}},\ and\
  \bibinfo {author} {\bibfnamefont {D.}~\bibnamefont {Sievenpiper}},\
  }\bibfield  {title} {\bibinfo {title} {Metasurfaces and their applications},\
  }\href {https://doi.org/doi:10.1515/nanoph-2017-0120} {\bibfield  {journal}
  {\bibinfo  {journal} {Nanophotonics}\ }\textbf {\bibinfo {volume} {7}},\
  \bibinfo {pages} {989} (\bibinfo {year} {2018})}\BibitemShut {NoStop}%
\bibitem [{\citenamefont {Wang}\ \emph {et~al.}(2018)\citenamefont {Wang},
  \citenamefont {Chernikov}, \citenamefont {Glazov}, \citenamefont {Heinz},
  \citenamefont {Marie}, \citenamefont {Amand},\ and\ \citenamefont
  {Urbaszek}}]{Wang_RevModPhys_2018}%
  \BibitemOpen
  \bibfield  {author} {\bibinfo {author} {\bibfnamefont {G.}~\bibnamefont
  {Wang}}, \bibinfo {author} {\bibfnamefont {A.}~\bibnamefont {Chernikov}},
  \bibinfo {author} {\bibfnamefont {M.~M.}\ \bibnamefont {Glazov}}, \bibinfo
  {author} {\bibfnamefont {T.~F.}\ \bibnamefont {Heinz}}, \bibinfo {author}
  {\bibfnamefont {X.}~\bibnamefont {Marie}}, \bibinfo {author} {\bibfnamefont
  {T.}~\bibnamefont {Amand}},\ and\ \bibinfo {author} {\bibfnamefont
  {B.}~\bibnamefont {Urbaszek}},\ }\bibfield  {title} {\bibinfo {title}
  {Colloquium: {Excitons} in atomically thin transition metal
  dichalcogenides},\ }\href {https://doi.org/10.1103/RevModPhys.90.021001}
  {\bibfield  {journal} {\bibinfo  {journal} {Rev. Mod. Phys.}\ }\textbf
  {\bibinfo {volume} {90}},\ \bibinfo {pages} {021001} (\bibinfo {year}
  {2018})}\BibitemShut {NoStop}%
\bibitem [{\citenamefont {Verre}\ \emph {et~al.}(2019)\citenamefont {Verre},
  \citenamefont {Baranov}, \citenamefont {Munkhbat}, \citenamefont {Cuadra},
  \citenamefont {K\"all},\ and\ \citenamefont {Shegai}}]{Verre_NatNano_2019}%
  \BibitemOpen
  \bibfield  {author} {\bibinfo {author} {\bibfnamefont {R.}~\bibnamefont
  {Verre}}, \bibinfo {author} {\bibfnamefont {D.~G.}\ \bibnamefont {Baranov}},
  \bibinfo {author} {\bibfnamefont {B.}~\bibnamefont {Munkhbat}}, \bibinfo
  {author} {\bibfnamefont {J.}~\bibnamefont {Cuadra}}, \bibinfo {author}
  {\bibfnamefont {M.}~\bibnamefont {K\"all}},\ and\ \bibinfo {author}
  {\bibfnamefont {T.}~\bibnamefont {Shegai}},\ }\bibfield  {title} {\bibinfo
  {title} {Transition metal dichalcogenide nanodisks as high-index dielectric
  {Mie} nanoresonators},\ }\href {https://doi.org/10.1038/s41565-019-0442-x}
  {\bibfield  {journal} {\bibinfo  {journal} {Nat. Nanotechnol.}\ }\textbf
  {\bibinfo {volume} {14}},\ \bibinfo {pages} {679} (\bibinfo {year}
  {2019})}\BibitemShut {NoStop}%
\bibitem [{\citenamefont {Munkhbat}\ \emph {et~al.}(2022)\citenamefont
  {Munkhbat}, \citenamefont {K\"u\c{c}\"uk\"oz}, \citenamefont {Baranov},
  \citenamefont {Antosiewicz},\ and\ \citenamefont
  {Shegai}}]{Munkhbat_lpor_2022}%
  \BibitemOpen
  \bibfield  {author} {\bibinfo {author} {\bibfnamefont {B.}~\bibnamefont
  {Munkhbat}}, \bibinfo {author} {\bibfnamefont {B.}~\bibnamefont
  {K\"u\c{c}\"uk\"oz}}, \bibinfo {author} {\bibfnamefont {D.~G.}\ \bibnamefont
  {Baranov}}, \bibinfo {author} {\bibfnamefont {T.~J.}\ \bibnamefont
  {Antosiewicz}},\ and\ \bibinfo {author} {\bibfnamefont {T.~O.}\ \bibnamefont
  {Shegai}},\ }\bibfield  {title} {\bibinfo {title} {Nanostructured transition
  metal dichalcogenide multilayers for advanced nanophotonics},\ }\href
  {https://doi.org/10.1002/lpor.202200057} {\bibfield  {journal} {\bibinfo
  {journal} {Laser Photonics Rev.}\ ,\ \bibinfo {pages} {2200057}} (\bibinfo
  {year} {2022})}\BibitemShut {NoStop}%
\bibitem [{\citenamefont {Xiao}\ \emph {et~al.}(2020)\citenamefont {Xiao},
  \citenamefont {Wang}, \citenamefont {Liu}, \citenamefont {Zhou},
  \citenamefont {Jiang},\ and\ \citenamefont {Zhang}}]{Xiao_2020}%
  \BibitemOpen
  \bibfield  {author} {\bibinfo {author} {\bibfnamefont {S.}~\bibnamefont
  {Xiao}}, \bibinfo {author} {\bibfnamefont {T.}~\bibnamefont {Wang}}, \bibinfo
  {author} {\bibfnamefont {T.}~\bibnamefont {Liu}}, \bibinfo {author}
  {\bibfnamefont {C.}~\bibnamefont {Zhou}}, \bibinfo {author} {\bibfnamefont
  {X.}~\bibnamefont {Jiang}},\ and\ \bibinfo {author} {\bibfnamefont
  {J.}~\bibnamefont {Zhang}},\ }\bibfield  {title} {\bibinfo {title} {Active
  metamaterials and metadevices: a review},\ }\href
  {https://doi.org/10.1088/1361-6463/abaced} {\bibfield  {journal} {\bibinfo
  {journal} {J. Phys. D: Appl. Phys.}\ }\textbf {\bibinfo {volume} {53}},\
  \bibinfo {pages} {503002} (\bibinfo {year} {2020})}\BibitemShut {NoStop}%
\bibitem [{\citenamefont {Yan}\ \emph {et~al.}(2015)\citenamefont {Yan},
  \citenamefont {Zhu}, \citenamefont {Zhou}, \citenamefont {E}, \citenamefont
  {Wang},\ and\ \citenamefont {Xu}}]{Yan_ApplOpt_2015}%
  \BibitemOpen
  \bibfield  {author} {\bibinfo {author} {\bibfnamefont {X.}~\bibnamefont
  {Yan}}, \bibinfo {author} {\bibfnamefont {L.}~\bibnamefont {Zhu}}, \bibinfo
  {author} {\bibfnamefont {Y.}~\bibnamefont {Zhou}}, \bibinfo {author}
  {\bibfnamefont {Y.}~\bibnamefont {E}}, \bibinfo {author} {\bibfnamefont
  {L.}~\bibnamefont {Wang}},\ and\ \bibinfo {author} {\bibfnamefont
  {X.}~\bibnamefont {Xu}},\ }\bibfield  {title} {\bibinfo {title} {Dielectric
  property of {MoS$_2$} crystal in terahertz and visible regions},\ }\href
  {https://doi.org/10.1364/AO.54.006732} {\bibfield  {journal} {\bibinfo
  {journal} {Appl. Opt.}\ }\textbf {\bibinfo {volume} {54}},\ \bibinfo {pages}
  {6732} (\bibinfo {year} {2015})}\BibitemShut {NoStop}%
\bibitem [{\citenamefont {Ermolaev}\ \emph {et~al.}(2020)\citenamefont
  {Ermolaev}, \citenamefont {Stebunov}, \citenamefont {Vyshnevyy},
  \citenamefont {Tatarkin}, \citenamefont {Yakubovsky}, \citenamefont
  {Novikov}, \citenamefont {Baranov}, \citenamefont {Shegai}, \citenamefont
  {Nikitin}, \citenamefont {Arsenin},\ and\ \citenamefont
  {Volkov}}]{Ermolaev_npj_2020}%
  \BibitemOpen
  \bibfield  {author} {\bibinfo {author} {\bibfnamefont {G.~A.}\ \bibnamefont
  {Ermolaev}}, \bibinfo {author} {\bibfnamefont {Y.~V.}\ \bibnamefont
  {Stebunov}}, \bibinfo {author} {\bibfnamefont {A.~A.}\ \bibnamefont
  {Vyshnevyy}}, \bibinfo {author} {\bibfnamefont {D.~E.}\ \bibnamefont
  {Tatarkin}}, \bibinfo {author} {\bibfnamefont {D.~I.}\ \bibnamefont
  {Yakubovsky}}, \bibinfo {author} {\bibfnamefont {S.~M.}\ \bibnamefont
  {Novikov}}, \bibinfo {author} {\bibfnamefont {D.~G.}\ \bibnamefont
  {Baranov}}, \bibinfo {author} {\bibfnamefont {T.}~\bibnamefont {Shegai}},
  \bibinfo {author} {\bibfnamefont {A.~Y.}\ \bibnamefont {Nikitin}}, \bibinfo
  {author} {\bibfnamefont {A.~V.}\ \bibnamefont {Arsenin}},\ and\ \bibinfo
  {author} {\bibfnamefont {V.~S.}\ \bibnamefont {Volkov}},\ }\bibfield  {title}
  {\bibinfo {title} {Broadband optical properties of monolayer and bulk
  {MoS$_2$}},\ }\href {https://doi.org/10.1038/s41699-020-0155-x} {\bibfield
  {journal} {\bibinfo  {journal} {npj 2D Mater. Appl.}\ }\textbf {\bibinfo
  {volume} {4}},\ \bibinfo {pages} {21} (\bibinfo {year} {2020})}\BibitemShut
  {NoStop}%
\bibitem [{\citenamefont {Islam}\ \emph {et~al.}(2021)\citenamefont {Islam},
  \citenamefont {Synowicki}, \citenamefont {Ismael}, \citenamefont {Oguntoye},
  \citenamefont {Grinalds},\ and\ \citenamefont {Escarra}}]{Islam_adpr_2021}%
  \BibitemOpen
  \bibfield  {author} {\bibinfo {author} {\bibfnamefont {K.~M.}\ \bibnamefont
  {Islam}}, \bibinfo {author} {\bibfnamefont {R.}~\bibnamefont {Synowicki}},
  \bibinfo {author} {\bibfnamefont {T.}~\bibnamefont {Ismael}}, \bibinfo
  {author} {\bibfnamefont {I.}~\bibnamefont {Oguntoye}}, \bibinfo {author}
  {\bibfnamefont {N.}~\bibnamefont {Grinalds}},\ and\ \bibinfo {author}
  {\bibfnamefont {M.~D.}\ \bibnamefont {Escarra}},\ }\bibfield  {title}
  {\bibinfo {title} {In-plane and out-of-plane optical properties of monolayer,
  few-layer, and thin-film {MoS$_2$} from 190 to 1700 nm and their application
  in photonic device design},\ }\href
  {https://doi.org/https://doi.org/10.1002/adpr.202000180} {\bibfield
  {journal} {\bibinfo  {journal} {Adv. Photonics Res.}\ }\textbf {\bibinfo
  {volume} {2}},\ \bibinfo {pages} {2000180} (\bibinfo {year}
  {2021})}\BibitemShut {NoStop}%
\bibitem [{\citenamefont {Ermolaev}\ \emph {et~al.}(2021)\citenamefont
  {Ermolaev}, \citenamefont {Grudinin}, \citenamefont {Stebunov}, \citenamefont
  {Voronin}, \citenamefont {Kravets}, \citenamefont {Duan}, \citenamefont
  {Mazitov}, \citenamefont {Tselikov}, \citenamefont {Bylinkin}, \citenamefont
  {Yakubovsky}, \citenamefont {Novikov}, \citenamefont {Baranov}, \citenamefont
  {Nikitin}, \citenamefont {Kruglov}, \citenamefont {Shegai}, \citenamefont
  {Alonso-Gonz\'alez}, \citenamefont {Grigorenko}, \citenamefont {Arsenin},
  \citenamefont {Novoselov},\ and\ \citenamefont {Volkov}}]{Ermolaev_Nat_2021}%
  \BibitemOpen
  \bibfield  {author} {\bibinfo {author} {\bibfnamefont {G.~A.}\ \bibnamefont
  {Ermolaev}}, \bibinfo {author} {\bibfnamefont {D.~V.}\ \bibnamefont
  {Grudinin}}, \bibinfo {author} {\bibfnamefont {Y.~V.}\ \bibnamefont
  {Stebunov}}, \bibinfo {author} {\bibfnamefont {K.~V.}\ \bibnamefont
  {Voronin}}, \bibinfo {author} {\bibfnamefont {V.~G.}\ \bibnamefont
  {Kravets}}, \bibinfo {author} {\bibfnamefont {J.}~\bibnamefont {Duan}},
  \bibinfo {author} {\bibfnamefont {A.~B.}\ \bibnamefont {Mazitov}}, \bibinfo
  {author} {\bibfnamefont {G.~I.}\ \bibnamefont {Tselikov}}, \bibinfo {author}
  {\bibfnamefont {A.}~\bibnamefont {Bylinkin}}, \bibinfo {author}
  {\bibfnamefont {D.~I.}\ \bibnamefont {Yakubovsky}}, \bibinfo {author}
  {\bibfnamefont {S.~M.}\ \bibnamefont {Novikov}}, \bibinfo {author}
  {\bibfnamefont {D.~G.}\ \bibnamefont {Baranov}}, \bibinfo {author}
  {\bibfnamefont {A.~Y.}\ \bibnamefont {Nikitin}}, \bibinfo {author}
  {\bibfnamefont {I.~A.}\ \bibnamefont {Kruglov}}, \bibinfo {author}
  {\bibfnamefont {T.}~\bibnamefont {Shegai}}, \bibinfo {author} {\bibfnamefont
  {P.}~\bibnamefont {Alonso-Gonz\'alez}}, \bibinfo {author} {\bibfnamefont
  {A.~N.}\ \bibnamefont {Grigorenko}}, \bibinfo {author} {\bibfnamefont
  {A.~V.}\ \bibnamefont {Arsenin}}, \bibinfo {author} {\bibfnamefont {K.~S.}\
  \bibnamefont {Novoselov}},\ and\ \bibinfo {author} {\bibfnamefont {V.~S.}\
  \bibnamefont {Volkov}},\ }\bibfield  {title} {\bibinfo {title} {Giant optical
  anisotropy in transition metal dichalcogenides for next-generation
  photonics},\ }\href {https://doi.org/10.1038/s41467-021-21139-x} {\bibfield
  {journal} {\bibinfo  {journal} {Nat. Commun.}\ }\textbf {\bibinfo {volume}
  {12}},\ \bibinfo {pages} {854} (\bibinfo {year} {2021})}\BibitemShut
  {NoStop}%
\bibitem [{\citenamefont {Born}\ and\ \citenamefont {Wolf}(1959)}]{born_book}%
  \BibitemOpen
  \bibfield  {author} {\bibinfo {author} {\bibfnamefont {M.}~\bibnamefont
  {Born}}\ and\ \bibinfo {author} {\bibfnamefont {E.}~\bibnamefont {Wolf}},\
  }\href@noop {} {\emph {\bibinfo {title} {Principles of Optics:
  Electromagnetic Theory of Propagation, Interference and Diffraction of
  Light}}}\ (\bibinfo  {publisher} {Pergamon Press Ltd.},\ \bibinfo {year}
  {1959})\ Chap.~\bibinfo {chapter} {14}\BibitemShut {NoStop}%
\bibitem [{\citenamefont {Yariv}\ and\ \citenamefont {Yeh}(1984)}]{yariv_book}%
  \BibitemOpen
  \bibfield  {author} {\bibinfo {author} {\bibfnamefont {A.}~\bibnamefont
  {Yariv}}\ and\ \bibinfo {author} {\bibfnamefont {P.}~\bibnamefont {Yeh}},\
  }\href@noop {} {\emph {\bibinfo {title} {Optical Waves in Crystals:
  Propagation and Control of Laser Radiation}}}\ (\bibinfo  {publisher}
  {Wiley},\ \bibinfo {year} {1984})\ Chap.~\bibinfo {chapter} {6}\BibitemShut
  {NoStop}%
\bibitem [{\citenamefont {Jayasekara}\ \emph {et~al.}(2016)\citenamefont
  {Jayasekara}, \citenamefont {Premaratne}, \citenamefont {Gunapala},\ and\
  \citenamefont {Stockman}}]{Stockman_JApplPhys_2016}%
  \BibitemOpen
  \bibfield  {author} {\bibinfo {author} {\bibfnamefont {C.}~\bibnamefont
  {Jayasekara}}, \bibinfo {author} {\bibfnamefont {M.}~\bibnamefont
  {Premaratne}}, \bibinfo {author} {\bibfnamefont {S.~D.}\ \bibnamefont
  {Gunapala}},\ and\ \bibinfo {author} {\bibfnamefont {M.~I.}\ \bibnamefont
  {Stockman}},\ }\bibfield  {title} {\bibinfo {title} {{MoS$_2$} spaser},\
  }\href {https://doi.org/10.1063/1.4945378} {\bibfield  {journal} {\bibinfo
  {journal} {J. Appl. Phys.}\ }\textbf {\bibinfo {volume} {119}},\ \bibinfo
  {pages} {133101} (\bibinfo {year} {2016})}\BibitemShut {NoStop}%
\bibitem [{\citenamefont {Ren}\ \emph {et~al.}(2021)\citenamefont {Ren},
  \citenamefont {Feng}, \citenamefont {Yao}, \citenamefont {Xu}, \citenamefont
  {Xin}, \citenamefont {Lan}, \citenamefont {You}, \citenamefont {Xiao},\ and\
  \citenamefont {Sha}}]{Ren_OptExpress_2021}%
  \BibitemOpen
  \bibfield  {author} {\bibinfo {author} {\bibfnamefont {Q.}~\bibnamefont
  {Ren}}, \bibinfo {author} {\bibfnamefont {F.}~\bibnamefont {Feng}}, \bibinfo
  {author} {\bibfnamefont {X.}~\bibnamefont {Yao}}, \bibinfo {author}
  {\bibfnamefont {Q.}~\bibnamefont {Xu}}, \bibinfo {author} {\bibfnamefont
  {M.}~\bibnamefont {Xin}}, \bibinfo {author} {\bibfnamefont {Z.}~\bibnamefont
  {Lan}}, \bibinfo {author} {\bibfnamefont {J.}~\bibnamefont {You}}, \bibinfo
  {author} {\bibfnamefont {X.}~\bibnamefont {Xiao}},\ and\ \bibinfo {author}
  {\bibfnamefont {W.~E.~I.}\ \bibnamefont {Sha}},\ }\bibfield  {title}
  {\bibinfo {title} {Multiplexing-oriented plasmon-{MoS2} hybrid metasurfaces
  driven by nonlinear quasi bound states in the continuum},\ }\href
  {https://doi.org/10.1364/OE.414730} {\bibfield  {journal} {\bibinfo
  {journal} {Opt. Express}\ }\textbf {\bibinfo {volume} {29}},\ \bibinfo
  {pages} {5384} (\bibinfo {year} {2021})}\BibitemShut {NoStop}%
\bibitem [{\citenamefont {Wang}\ \emph {et~al.}(2022)\citenamefont {Wang},
  \citenamefont {Xin}, \citenamefont {Ren}, \citenamefont {Cai}, \citenamefont
  {Han}, \citenamefont {Tian}, \citenamefont {Zhang}, \citenamefont {Jiang},
  \citenamefont {Lan}, \citenamefont {You},\ and\ \citenamefont
  {Sha}}]{Wang_ChinOptLett_2022}%
  \BibitemOpen
  \bibfield  {author} {\bibinfo {author} {\bibfnamefont {X.}~\bibnamefont
  {Wang}}, \bibinfo {author} {\bibfnamefont {J.}~\bibnamefont {Xin}}, \bibinfo
  {author} {\bibfnamefont {Q.}~\bibnamefont {Ren}}, \bibinfo {author}
  {\bibfnamefont {H.}~\bibnamefont {Cai}}, \bibinfo {author} {\bibfnamefont
  {J.}~\bibnamefont {Han}}, \bibinfo {author} {\bibfnamefont {C.}~\bibnamefont
  {Tian}}, \bibinfo {author} {\bibfnamefont {P.}~\bibnamefont {Zhang}},
  \bibinfo {author} {\bibfnamefont {L.}~\bibnamefont {Jiang}}, \bibinfo
  {author} {\bibfnamefont {Z.}~\bibnamefont {Lan}}, \bibinfo {author}
  {\bibfnamefont {J.}~\bibnamefont {You}},\ and\ \bibinfo {author}
  {\bibfnamefont {W.~E.~I.}\ \bibnamefont {Sha}},\ }\bibfield  {title}
  {\bibinfo {title} {Plasmon hybridization induced by quasi bound state in the
  continuum of graphene metasurfaces oriented for high-accuracy
  polarization-insensitive two-dimensional sensors},\ }\href
  {https://doi.org/10.1364/COL.20.042201} {\bibfield  {journal} {\bibinfo
  {journal} {Chin. Opt. Lett.}\ }\textbf {\bibinfo {volume} {20}},\ \bibinfo
  {pages} {042201} (\bibinfo {year} {2022})}\BibitemShut {NoStop}%
\bibitem [{\citenamefont {Srivastava}\ \emph {et~al.}(2017)\citenamefont
  {Srivastava}, \citenamefont {Chaturvedi}, \citenamefont {Manjappa},
  \citenamefont {Kumar}, \citenamefont {Dayal}, \citenamefont {Kloc},\ and\
  \citenamefont {Singh}}]{Srivastava_AdvOptMater_2017}%
  \BibitemOpen
  \bibfield  {author} {\bibinfo {author} {\bibfnamefont {Y.~K.}\ \bibnamefont
  {Srivastava}}, \bibinfo {author} {\bibfnamefont {A.}~\bibnamefont
  {Chaturvedi}}, \bibinfo {author} {\bibfnamefont {M.}~\bibnamefont
  {Manjappa}}, \bibinfo {author} {\bibfnamefont {A.}~\bibnamefont {Kumar}},
  \bibinfo {author} {\bibfnamefont {G.}~\bibnamefont {Dayal}}, \bibinfo
  {author} {\bibfnamefont {C.}~\bibnamefont {Kloc}},\ and\ \bibinfo {author}
  {\bibfnamefont {R.}~\bibnamefont {Singh}},\ }\bibfield  {title} {\bibinfo
  {title} {{MoS$_2$} for ultrafast all-optical switching and modulation of
  {THz} {Fano} metaphotonic devices},\ }\href
  {https://doi.org/https://doi.org/10.1002/adom.201700762} {\bibfield
  {journal} {\bibinfo  {journal} {Adv. Opt. Mater.}\ }\textbf {\bibinfo
  {volume} {5}},\ \bibinfo {pages} {1700762} (\bibinfo {year}
  {2017})}\BibitemShut {NoStop}%
\bibitem [{\citenamefont {Bucher}\ \emph {et~al.}(2019)\citenamefont {Bucher},
  \citenamefont {Vaskin}, \citenamefont {Mupparapu}, \citenamefont {L\"ochner},
  \citenamefont {George}, \citenamefont {Chong}, \citenamefont {Fasold},
  \citenamefont {Neumann}, \citenamefont {Choi}, \citenamefont {Eilenberger},
  \citenamefont {Setzpfandt}, \citenamefont {Kivshar}, \citenamefont {Pertsch},
  \citenamefont {Turchanin},\ and\ \citenamefont
  {Staude}}]{Bucher_ACSPhotonics_2019}%
  \BibitemOpen
  \bibfield  {author} {\bibinfo {author} {\bibfnamefont {T.}~\bibnamefont
  {Bucher}}, \bibinfo {author} {\bibfnamefont {A.}~\bibnamefont {Vaskin}},
  \bibinfo {author} {\bibfnamefont {R.}~\bibnamefont {Mupparapu}}, \bibinfo
  {author} {\bibfnamefont {F.~J.~F.}\ \bibnamefont {L\"ochner}}, \bibinfo
  {author} {\bibfnamefont {A.}~\bibnamefont {George}}, \bibinfo {author}
  {\bibfnamefont {K.~E.}\ \bibnamefont {Chong}}, \bibinfo {author}
  {\bibfnamefont {S.}~\bibnamefont {Fasold}}, \bibinfo {author} {\bibfnamefont
  {C.}~\bibnamefont {Neumann}}, \bibinfo {author} {\bibfnamefont {D.-Y.}\
  \bibnamefont {Choi}}, \bibinfo {author} {\bibfnamefont {F.}~\bibnamefont
  {Eilenberger}}, \bibinfo {author} {\bibfnamefont {F.}~\bibnamefont
  {Setzpfandt}}, \bibinfo {author} {\bibfnamefont {Y.~S.}\ \bibnamefont
  {Kivshar}}, \bibinfo {author} {\bibfnamefont {T.}~\bibnamefont {Pertsch}},
  \bibinfo {author} {\bibfnamefont {A.}~\bibnamefont {Turchanin}},\ and\
  \bibinfo {author} {\bibfnamefont {I.}~\bibnamefont {Staude}},\ }\bibfield
  {title} {\bibinfo {title} {Tailoring photoluminescence from {MoS$_2$}
  monolayers by {Mie}-resonant metasurfaces},\ }\href
  {https://doi.org/10.1021/acsphotonics.8b01771} {\bibfield  {journal}
  {\bibinfo  {journal} {ACS Photonics}\ }\textbf {\bibinfo {volume} {6}},\
  \bibinfo {pages} {1002} (\bibinfo {year} {2019})}\BibitemShut {NoStop}%
\bibitem [{\citenamefont {Muhammad}\ \emph {et~al.}(2021)\citenamefont
  {Muhammad}, \citenamefont {Chen}, \citenamefont {Qiu},\ and\ \citenamefont
  {Wang}}]{Muhammad_NanoLett_2021}%
  \BibitemOpen
  \bibfield  {author} {\bibinfo {author} {\bibfnamefont {N.}~\bibnamefont
  {Muhammad}}, \bibinfo {author} {\bibfnamefont {Y.}~\bibnamefont {Chen}},
  \bibinfo {author} {\bibfnamefont {C.-W.}\ \bibnamefont {Qiu}},\ and\ \bibinfo
  {author} {\bibfnamefont {G.~P.}\ \bibnamefont {Wang}},\ }\bibfield  {title}
  {\bibinfo {title} {Optical bound states in continuum in {MoS$_2$}-based
  metasurface for directional light emission},\ }\href
  {https://doi.org/10.1021/acs.nanolett.0c03818} {\bibfield  {journal}
  {\bibinfo  {journal} {Nano Lett.}\ }\textbf {\bibinfo {volume} {21}},\
  \bibinfo {pages} {967} (\bibinfo {year} {2021})}\BibitemShut {NoStop}%
\bibitem [{\citenamefont {Li}\ \emph {et~al.}(2022)\citenamefont {Li},
  \citenamefont {Wei}, \citenamefont {Zhou}, \citenamefont {Xiao},
  \citenamefont {Qin}, \citenamefont {Xia},\ and\ \citenamefont
  {Wu}}]{Li_PhysRevB_2022}%
  \BibitemOpen
  \bibfield  {author} {\bibinfo {author} {\bibfnamefont {H.}~\bibnamefont
  {Li}}, \bibinfo {author} {\bibfnamefont {G.}~\bibnamefont {Wei}}, \bibinfo
  {author} {\bibfnamefont {H.}~\bibnamefont {Zhou}}, \bibinfo {author}
  {\bibfnamefont {H.}~\bibnamefont {Xiao}}, \bibinfo {author} {\bibfnamefont
  {M.}~\bibnamefont {Qin}}, \bibinfo {author} {\bibfnamefont {S.}~\bibnamefont
  {Xia}},\ and\ \bibinfo {author} {\bibfnamefont {F.}~\bibnamefont {Wu}},\
  }\bibfield  {title} {\bibinfo {title} {Polarization-independent near-infrared
  superabsorption in transition metal dichalcogenide {Huygens} metasurfaces by
  degenerate critical coupling},\ }\href
  {https://doi.org/10.1103/PhysRevB.105.165305} {\bibfield  {journal} {\bibinfo
   {journal} {Phys. Rev. B}\ }\textbf {\bibinfo {volume} {105}},\ \bibinfo
  {pages} {165305} (\bibinfo {year} {2022})}\BibitemShut {NoStop}%
\bibitem [{\citenamefont {Zhang}\ \emph {et~al.}(2022)\citenamefont {Zhang},
  \citenamefont {Chen}, \citenamefont {Ma}, \citenamefont {You}, \citenamefont
  {Zhang}, \citenamefont {Fan},\ and\ \citenamefont
  {Zhou}}]{Zhang_OptLett_2022}%
  \BibitemOpen
  \bibfield  {author} {\bibinfo {author} {\bibfnamefont {Y.}~\bibnamefont
  {Zhang}}, \bibinfo {author} {\bibfnamefont {D.}~\bibnamefont {Chen}},
  \bibinfo {author} {\bibfnamefont {W.}~\bibnamefont {Ma}}, \bibinfo {author}
  {\bibfnamefont {S.}~\bibnamefont {You}}, \bibinfo {author} {\bibfnamefont
  {J.}~\bibnamefont {Zhang}}, \bibinfo {author} {\bibfnamefont
  {M.}~\bibnamefont {Fan}},\ and\ \bibinfo {author} {\bibfnamefont
  {C.}~\bibnamefont {Zhou}},\ }\bibfield  {title} {\bibinfo {title} {Active
  optical modulation of quasi-{BICs} in {Si--VO$_2$} hybrid metasurfaces},\
  }\href {https://doi.org/10.1364/OL.472927} {\bibfield  {journal} {\bibinfo
  {journal} {Opt. Lett.}\ }\textbf {\bibinfo {volume} {47}},\ \bibinfo {pages}
  {5517} (\bibinfo {year} {2022})}\BibitemShut {NoStop}%
\bibitem [{\citenamefont {Evlyukhin}\ \emph {et~al.}(2010)\citenamefont
  {Evlyukhin}, \citenamefont {Reinhardt}, \citenamefont {Seidel}, \citenamefont
  {Luk'yanchuk},\ and\ \citenamefont {Chichkov}}]{evlyukhin2010optical}%
  \BibitemOpen
  \bibfield  {author} {\bibinfo {author} {\bibfnamefont {A.~B.}\ \bibnamefont
  {Evlyukhin}}, \bibinfo {author} {\bibfnamefont {C.}~\bibnamefont
  {Reinhardt}}, \bibinfo {author} {\bibfnamefont {A.}~\bibnamefont {Seidel}},
  \bibinfo {author} {\bibfnamefont {B.~S.}\ \bibnamefont {Luk'yanchuk}},\ and\
  \bibinfo {author} {\bibfnamefont {B.~N.}\ \bibnamefont {Chichkov}},\
  }\bibfield  {title} {\bibinfo {title} {Optical response features of
  {Si}-nanoparticle arrays},\ }\href
  {https://doi.org/10.1103/PhysRevB.82.045404} {\bibfield  {journal} {\bibinfo
  {journal} {Phys. Rev. B}\ }\textbf {\bibinfo {volume} {82}},\ \bibinfo
  {pages} {045404} (\bibinfo {year} {2010})}\BibitemShut {NoStop}%
\bibitem [{\citenamefont {Kuznetsov}\ \emph {et~al.}(2016)\citenamefont
  {Kuznetsov}, \citenamefont {Miroshnichenko}, \citenamefont {Brongersma},
  \citenamefont {Kivshar},\ and\ \citenamefont
  {Luk’yanchuk}}]{Kuznetsov_Science_2016}%
  \BibitemOpen
  \bibfield  {author} {\bibinfo {author} {\bibfnamefont {A.~I.}\ \bibnamefont
  {Kuznetsov}}, \bibinfo {author} {\bibfnamefont {A.~E.}\ \bibnamefont
  {Miroshnichenko}}, \bibinfo {author} {\bibfnamefont {M.~L.}\ \bibnamefont
  {Brongersma}}, \bibinfo {author} {\bibfnamefont {Y.~S.}\ \bibnamefont
  {Kivshar}},\ and\ \bibinfo {author} {\bibfnamefont {B.}~\bibnamefont
  {Luk’yanchuk}},\ }\bibfield  {title} {\bibinfo {title} {Optically resonant
  dielectric nanostructures},\ }\href {https://doi.org/10.1126/science.aag2472}
  {\bibfield  {journal} {\bibinfo  {journal} {Science}\ }\textbf {\bibinfo
  {volume} {354}},\ \bibinfo {pages} {aag2472} (\bibinfo {year}
  {2016})}\BibitemShut {NoStop}%
\bibitem [{\citenamefont {Evlyukhin}\ \emph {et~al.}(2011)\citenamefont
  {Evlyukhin}, \citenamefont {Reinhardt},\ and\ \citenamefont
  {Chichkov}}]{evlyukhin2011multipole}%
  \BibitemOpen
  \bibfield  {author} {\bibinfo {author} {\bibfnamefont {A.~B.}\ \bibnamefont
  {Evlyukhin}}, \bibinfo {author} {\bibfnamefont {C.}~\bibnamefont
  {Reinhardt}},\ and\ \bibinfo {author} {\bibfnamefont {B.~N.}\ \bibnamefont
  {Chichkov}},\ }\bibfield  {title} {\bibinfo {title} {Multipole light
  scattering by nonspherical nanoparticles in the discrete dipole
  approximation},\ }\href {https://doi.org/10.1103/PhysRevB.84.235429}
  {\bibfield  {journal} {\bibinfo  {journal} {Phys. Rev. B}\ }\textbf {\bibinfo
  {volume} {84}},\ \bibinfo {pages} {235429} (\bibinfo {year}
  {2011})}\BibitemShut {NoStop}%
\bibitem [{\citenamefont {Khromova}\ \emph {et~al.}(2016)\citenamefont
  {Khromova}, \citenamefont {Ku\v{z}el}, \citenamefont {Brener}, \citenamefont
  {Reno}, \citenamefont {{Chung Seu}}, \citenamefont {Elissalde}, \citenamefont
  {Maglione}, \citenamefont {Mounaix},\ and\ \citenamefont
  {Mitrofanov}}]{Khromova_Laser_2016}%
  \BibitemOpen
  \bibfield  {author} {\bibinfo {author} {\bibfnamefont {I.}~\bibnamefont
  {Khromova}}, \bibinfo {author} {\bibfnamefont {P.}~\bibnamefont {Ku\v{z}el}},
  \bibinfo {author} {\bibfnamefont {I.}~\bibnamefont {Brener}}, \bibinfo
  {author} {\bibfnamefont {J.~L.}\ \bibnamefont {Reno}}, \bibinfo {author}
  {\bibfnamefont {U.-C.}\ \bibnamefont {{Chung Seu}}}, \bibinfo {author}
  {\bibfnamefont {C.}~\bibnamefont {Elissalde}}, \bibinfo {author}
  {\bibfnamefont {M.}~\bibnamefont {Maglione}}, \bibinfo {author}
  {\bibfnamefont {P.}~\bibnamefont {Mounaix}},\ and\ \bibinfo {author}
  {\bibfnamefont {O.}~\bibnamefont {Mitrofanov}},\ }\bibfield  {title}
  {\bibinfo {title} {Splitting of magnetic dipole modes in anisotropic
  {TiO$_2$} micro-spheres},\ }\href {https://doi.org/10.1002/lpor.201600084}
  {\bibfield  {journal} {\bibinfo  {journal} {Laser Photon. Rev.}\ }\textbf
  {\bibinfo {volume} {10}},\ \bibinfo {pages} {681} (\bibinfo {year}
  {2016})}\BibitemShut {NoStop}%
\bibitem [{\citenamefont {Razmjooei}\ \emph {et~al.}(2021)\citenamefont
  {Razmjooei}, \citenamefont {Ko}, \citenamefont {Simlan},\ and\ \citenamefont
  {Magnusson}}]{Razmjooei_OptExpress_2021}%
  \BibitemOpen
  \bibfield  {author} {\bibinfo {author} {\bibfnamefont {N.}~\bibnamefont
  {Razmjooei}}, \bibinfo {author} {\bibfnamefont {Y.~H.}\ \bibnamefont {Ko}},
  \bibinfo {author} {\bibfnamefont {F.~A.}\ \bibnamefont {Simlan}},\ and\
  \bibinfo {author} {\bibfnamefont {R.}~\bibnamefont {Magnusson}},\ }\bibfield
  {title} {\bibinfo {title} {Resonant reflection by microsphere arrays with
  {AR}-quenched {Mie} scattering},\ }\href {https://doi.org/10.1364/OE.427982}
  {\bibfield  {journal} {\bibinfo  {journal} {Opt. Express}\ }\textbf {\bibinfo
  {volume} {29}},\ \bibinfo {pages} {19183} (\bibinfo {year}
  {2021})}\BibitemShut {NoStop}%
\bibitem [{\citenamefont {Decker}\ \emph {et~al.}(2015)\citenamefont {Decker},
  \citenamefont {Staude}, \citenamefont {Falkner}, \citenamefont {Dominguez},
  \citenamefont {Neshev}, \citenamefont {Brener}, \citenamefont {Pertsch},\
  and\ \citenamefont {Kivshar}}]{Decker_AdvOptMat_2015}%
  \BibitemOpen
  \bibfield  {author} {\bibinfo {author} {\bibfnamefont {M.}~\bibnamefont
  {Decker}}, \bibinfo {author} {\bibfnamefont {I.}~\bibnamefont {Staude}},
  \bibinfo {author} {\bibfnamefont {M.}~\bibnamefont {Falkner}}, \bibinfo
  {author} {\bibfnamefont {J.}~\bibnamefont {Dominguez}}, \bibinfo {author}
  {\bibfnamefont {D.~N.}\ \bibnamefont {Neshev}}, \bibinfo {author}
  {\bibfnamefont {I.}~\bibnamefont {Brener}}, \bibinfo {author} {\bibfnamefont
  {T.}~\bibnamefont {Pertsch}},\ and\ \bibinfo {author} {\bibfnamefont {Y.~S.}\
  \bibnamefont {Kivshar}},\ }\bibfield  {title} {\bibinfo {title}
  {High-efficiency dielectric {Huygens'} surfaces},\ }\href
  {https://doi.org/https://doi.org/10.1002/adom.201400584} {\bibfield
  {journal} {\bibinfo  {journal} {Adv. Opt. Mater.}\ }\textbf {\bibinfo
  {volume} {3}},\ \bibinfo {pages} {813} (\bibinfo {year} {2015})}\BibitemShut
  {NoStop}%
\bibitem [{\citenamefont {Fedotov}\ \emph {et~al.}(2007)\citenamefont
  {Fedotov}, \citenamefont {Rose}, \citenamefont {Prosvirnin}, \citenamefont
  {Papasimakis},\ and\ \citenamefont {Zheludev}}]{fedotov2007sharp}%
  \BibitemOpen
  \bibfield  {author} {\bibinfo {author} {\bibfnamefont {V.~A.}\ \bibnamefont
  {Fedotov}}, \bibinfo {author} {\bibfnamefont {M.}~\bibnamefont {Rose}},
  \bibinfo {author} {\bibfnamefont {S.~L.}\ \bibnamefont {Prosvirnin}},
  \bibinfo {author} {\bibfnamefont {N.}~\bibnamefont {Papasimakis}},\ and\
  \bibinfo {author} {\bibfnamefont {N.~I.}\ \bibnamefont {Zheludev}},\
  }\bibfield  {title} {\bibinfo {title} {Sharp trapped-mode resonances in
  planar metamaterials with a broken structural symmetry},\ }\href
  {https://doi.org/10.1103/PhysRevLett.99.147401} {\bibfield  {journal}
  {\bibinfo  {journal} {Phys. Rev. Lett.}\ }\textbf {\bibinfo {volume} {99}},\
  \bibinfo {pages} {147401} (\bibinfo {year} {2007})}\BibitemShut {NoStop}%
\bibitem [{\citenamefont {Koshelev}\ \emph {et~al.}(2018)\citenamefont
  {Koshelev}, \citenamefont {Lepeshov}, \citenamefont {Liu}, \citenamefont
  {Bogdanov},\ and\ \citenamefont {Kivshar}}]{Koshelev_PhysRevLett_2018}%
  \BibitemOpen
  \bibfield  {author} {\bibinfo {author} {\bibfnamefont {K.}~\bibnamefont
  {Koshelev}}, \bibinfo {author} {\bibfnamefont {S.}~\bibnamefont {Lepeshov}},
  \bibinfo {author} {\bibfnamefont {M.}~\bibnamefont {Liu}}, \bibinfo {author}
  {\bibfnamefont {A.}~\bibnamefont {Bogdanov}},\ and\ \bibinfo {author}
  {\bibfnamefont {Y.}~\bibnamefont {Kivshar}},\ }\bibfield  {title} {\bibinfo
  {title} {Asymmetric metasurfaces with high-{$Q$} resonances governed by bound
  states in the continuum},\ }\href
  {https://doi.org/10.1103/PhysRevLett.121.193903} {\bibfield  {journal}
  {\bibinfo  {journal} {Phys. Rev. Lett.}\ }\textbf {\bibinfo {volume} {121}},\
  \bibinfo {pages} {193903} (\bibinfo {year} {2018})}\BibitemShut {NoStop}%
\bibitem [{\citenamefont {Khardikov}\ \emph {et~al.}(2010)\citenamefont
  {Khardikov}, \citenamefont {Iarko},\ and\ \citenamefont
  {Prosvirnin}}]{Khardikov_2010}%
  \BibitemOpen
  \bibfield  {author} {\bibinfo {author} {\bibfnamefont {V.~V.}\ \bibnamefont
  {Khardikov}}, \bibinfo {author} {\bibfnamefont {E.~O.}\ \bibnamefont
  {Iarko}},\ and\ \bibinfo {author} {\bibfnamefont {S.~L.}\ \bibnamefont
  {Prosvirnin}},\ }\bibfield  {title} {\bibinfo {title} {Trapping of light by
  metal arrays},\ }\href {https://doi.org/10.1088/2040-8978/12/4/045102}
  {\bibfield  {journal} {\bibinfo  {journal} {J. Opt.}\ }\textbf {\bibinfo
  {volume} {12}},\ \bibinfo {pages} {045102} (\bibinfo {year}
  {2010})}\BibitemShut {NoStop}%
\bibitem [{\citenamefont {Kupriianov}\ \emph {et~al.}(2019)\citenamefont
  {Kupriianov}, \citenamefont {Xu}, \citenamefont {Sayanskiy}, \citenamefont
  {Dmitriev}, \citenamefont {Kivshar},\ and\ \citenamefont
  {Tuz}}]{Kupriianov_PhysRevApplied_2019}%
  \BibitemOpen
  \bibfield  {author} {\bibinfo {author} {\bibfnamefont {A.~S.}\ \bibnamefont
  {Kupriianov}}, \bibinfo {author} {\bibfnamefont {Y.}~\bibnamefont {Xu}},
  \bibinfo {author} {\bibfnamefont {A.}~\bibnamefont {Sayanskiy}}, \bibinfo
  {author} {\bibfnamefont {V.}~\bibnamefont {Dmitriev}}, \bibinfo {author}
  {\bibfnamefont {Y.~S.}\ \bibnamefont {Kivshar}},\ and\ \bibinfo {author}
  {\bibfnamefont {V.~R.}\ \bibnamefont {Tuz}},\ }\bibfield  {title} {\bibinfo
  {title} {Metasurface engineering through bound states in the continuum},\
  }\href {https://doi.org/10.1103/PhysRevApplied.12.014024} {\bibfield
  {journal} {\bibinfo  {journal} {Phys. Rev. Applied}\ }\textbf {\bibinfo
  {volume} {12}},\ \bibinfo {pages} {014024} (\bibinfo {year}
  {2019})}\BibitemShut {NoStop}%
\bibitem [{\citenamefont {Xiao}\ \emph {et~al.}(2022)\citenamefont {Xiao},
  \citenamefont {Qin}, \citenamefont {Duan}, \citenamefont {Wu},\ and\
  \citenamefont {Liu}}]{Xiao_PhysRevB_2022}%
  \BibitemOpen
  \bibfield  {author} {\bibinfo {author} {\bibfnamefont {S.}~\bibnamefont
  {Xiao}}, \bibinfo {author} {\bibfnamefont {M.}~\bibnamefont {Qin}}, \bibinfo
  {author} {\bibfnamefont {J.}~\bibnamefont {Duan}}, \bibinfo {author}
  {\bibfnamefont {F.}~\bibnamefont {Wu}},\ and\ \bibinfo {author}
  {\bibfnamefont {T.}~\bibnamefont {Liu}},\ }\bibfield  {title} {\bibinfo
  {title} {Polarization-controlled dynamically switchable high-harmonic
  generation from all-dielectric metasurfaces governed by dual bound states in
  the continuum},\ }\href {https://doi.org/10.1103/PhysRevB.105.195440}
  {\bibfield  {journal} {\bibinfo  {journal} {Phys. Rev. B}\ }\textbf {\bibinfo
  {volume} {105}},\ \bibinfo {pages} {195440} (\bibinfo {year}
  {2022})}\BibitemShut {NoStop}%
\bibitem [{\citenamefont {He}\ \emph {et~al.}(2018)\citenamefont {He},
  \citenamefont {Guo}, \citenamefont {Feng}, \citenamefont {Xu},\ and\
  \citenamefont {Miroshnichenko}}]{He_PhysRevB_2018}%
  \BibitemOpen
  \bibfield  {author} {\bibinfo {author} {\bibfnamefont {Y.}~\bibnamefont
  {He}}, \bibinfo {author} {\bibfnamefont {G.}~\bibnamefont {Guo}}, \bibinfo
  {author} {\bibfnamefont {T.}~\bibnamefont {Feng}}, \bibinfo {author}
  {\bibfnamefont {Y.}~\bibnamefont {Xu}},\ and\ \bibinfo {author}
  {\bibfnamefont {A.~E.}\ \bibnamefont {Miroshnichenko}},\ }\bibfield  {title}
  {\bibinfo {title} {Toroidal dipole bound states in the continuum},\ }\href
  {https://doi.org/10.1103/PhysRevB.98.161112} {\bibfield  {journal} {\bibinfo
  {journal} {Phys. Rev. B}\ }\textbf {\bibinfo {volume} {98}},\ \bibinfo
  {pages} {161112} (\bibinfo {year} {2018})}\BibitemShut {NoStop}%
\bibitem [{\citenamefont {van Hoof}\ \emph {et~al.}(2021)\citenamefont {van
  Hoof}, \citenamefont {Abujetas}, \citenamefont {Ter~Huurne}, \citenamefont
  {Verdelli}, \citenamefont {Timmermans}, \citenamefont {S{\'a}nchez-Gil},\
  and\ \citenamefont {Rivas}}]{van2021unveiling}%
  \BibitemOpen
  \bibfield  {author} {\bibinfo {author} {\bibfnamefont {N.~J.}\ \bibnamefont
  {van Hoof}}, \bibinfo {author} {\bibfnamefont {D.~R.}\ \bibnamefont
  {Abujetas}}, \bibinfo {author} {\bibfnamefont {S.~E.}\ \bibnamefont
  {Ter~Huurne}}, \bibinfo {author} {\bibfnamefont {F.}~\bibnamefont
  {Verdelli}}, \bibinfo {author} {\bibfnamefont {G.~C.}\ \bibnamefont
  {Timmermans}}, \bibinfo {author} {\bibfnamefont {J.~A.}\ \bibnamefont
  {S{\'a}nchez-Gil}},\ and\ \bibinfo {author} {\bibfnamefont {J.~G.}\
  \bibnamefont {Rivas}},\ }\bibfield  {title} {\bibinfo {title} {Unveiling the
  symmetry protection of bound states in the continuum with terahertz
  near-field imaging},\ }\href {https://doi.org/10.1021/acsphotonics.1c00937}
  {\bibfield  {journal} {\bibinfo  {journal} {ACS Photonics}\ }\textbf
  {\bibinfo {volume} {8}},\ \bibinfo {pages} {3010} (\bibinfo {year}
  {2021})}\BibitemShut {NoStop}%
\bibitem [{\citenamefont {Evlyukhin}\ \emph {et~al.}(2020)\citenamefont
  {Evlyukhin}, \citenamefont {Tuz}, \citenamefont {Volkov},\ and\ \citenamefont
  {Chichkov}}]{Evlyukhin_PhysRevB_2020}%
  \BibitemOpen
  \bibfield  {author} {\bibinfo {author} {\bibfnamefont {A.~B.}\ \bibnamefont
  {Evlyukhin}}, \bibinfo {author} {\bibfnamefont {V.~R.}\ \bibnamefont {Tuz}},
  \bibinfo {author} {\bibfnamefont {V.~S.}\ \bibnamefont {Volkov}},\ and\
  \bibinfo {author} {\bibfnamefont {B.~N.}\ \bibnamefont {Chichkov}},\
  }\bibfield  {title} {\bibinfo {title} {Bianisotropy for light trapping in
  all-dielectric metasurfaces},\ }\href
  {https://doi.org/10.1103/PhysRevB.101.205415} {\bibfield  {journal} {\bibinfo
   {journal} {Phys. Rev. B}\ }\textbf {\bibinfo {volume} {101}},\ \bibinfo
  {pages} {205415} (\bibinfo {year} {2020})}\BibitemShut {NoStop}%
\bibitem [{\citenamefont {Tsilipakos}\ \emph {et~al.}(2021)\citenamefont
  {Tsilipakos}, \citenamefont {Maiolo}, \citenamefont {Maita}, \citenamefont
  {Beccherelli}, \citenamefont {Kafesaki}, \citenamefont {Kriezis},
  \citenamefont {Yioultsis},\ and\ \citenamefont
  {Zografopoulos}}]{Kafesaki_apl_2021}%
  \BibitemOpen
  \bibfield  {author} {\bibinfo {author} {\bibfnamefont {O.}~\bibnamefont
  {Tsilipakos}}, \bibinfo {author} {\bibfnamefont {L.}~\bibnamefont {Maiolo}},
  \bibinfo {author} {\bibfnamefont {F.}~\bibnamefont {Maita}}, \bibinfo
  {author} {\bibfnamefont {R.}~\bibnamefont {Beccherelli}}, \bibinfo {author}
  {\bibfnamefont {M.}~\bibnamefont {Kafesaki}}, \bibinfo {author}
  {\bibfnamefont {E.~E.}\ \bibnamefont {Kriezis}}, \bibinfo {author}
  {\bibfnamefont {T.~V.}\ \bibnamefont {Yioultsis}},\ and\ \bibinfo {author}
  {\bibfnamefont {D.~C.}\ \bibnamefont {Zografopoulos}},\ }\bibfield  {title}
  {\bibinfo {title} {Experimental demonstration of ultrathin broken-symmetry
  metasurfaces with controllably sharp resonant response},\ }\href
  {https://doi.org/10.1063/5.0073803} {\bibfield  {journal} {\bibinfo
  {journal} {Appl. Phys. Lett.}\ }\textbf {\bibinfo {volume} {119}},\ \bibinfo
  {pages} {231601} (\bibinfo {year} {2021})}\BibitemShut {NoStop}%
\bibitem [{\citenamefont {Tuz}\ \emph {et~al.}(2018)\citenamefont {Tuz},
  \citenamefont {Khardikov}, \citenamefont {Kupriianov}, \citenamefont
  {Domina}, \citenamefont {Xu}, \citenamefont {Wang},\ and\ \citenamefont
  {Sun}}]{Tuz_OptExpress_2018}%
  \BibitemOpen
  \bibfield  {author} {\bibinfo {author} {\bibfnamefont {V.~R.}\ \bibnamefont
  {Tuz}}, \bibinfo {author} {\bibfnamefont {V.~V.}\ \bibnamefont {Khardikov}},
  \bibinfo {author} {\bibfnamefont {A.~S.}\ \bibnamefont {Kupriianov}},
  \bibinfo {author} {\bibfnamefont {K.~L.}\ \bibnamefont {Domina}}, \bibinfo
  {author} {\bibfnamefont {S.}~\bibnamefont {Xu}}, \bibinfo {author}
  {\bibfnamefont {H.}~\bibnamefont {Wang}},\ and\ \bibinfo {author}
  {\bibfnamefont {H.-B.}\ \bibnamefont {Sun}},\ }\bibfield  {title} {\bibinfo
  {title} {High-quality trapped modes in all-dielectric metamaterials},\ }\href
  {https://doi.org/10.1364/OE.26.002905} {\bibfield  {journal} {\bibinfo
  {journal} {Opt. Express}\ }\textbf {\bibinfo {volume} {26}},\ \bibinfo
  {pages} {2905} (\bibinfo {year} {2018})}\BibitemShut {NoStop}%
\bibitem [{\citenamefont {Yu}\ \emph {et~al.}(2019)\citenamefont {Yu},
  \citenamefont {Kupriianov}, \citenamefont {Dmitriev},\ and\ \citenamefont
  {Tuz}}]{Tuz_JApplPhys_2019}%
  \BibitemOpen
  \bibfield  {author} {\bibinfo {author} {\bibfnamefont {P.}~\bibnamefont
  {Yu}}, \bibinfo {author} {\bibfnamefont {A.~S.}\ \bibnamefont {Kupriianov}},
  \bibinfo {author} {\bibfnamefont {V.}~\bibnamefont {Dmitriev}},\ and\
  \bibinfo {author} {\bibfnamefont {V.~R.}\ \bibnamefont {Tuz}},\ }\bibfield
  {title} {\bibinfo {title} {All-dielectric metasurfaces with trapped modes:
  Group-theoretical description},\ }\href {https://doi.org/10.1063/1.5087054}
  {\bibfield  {journal} {\bibinfo  {journal} {J. Appl. Phys.}\ }\textbf
  {\bibinfo {volume} {125}},\ \bibinfo {pages} {143101} (\bibinfo {year}
  {2019})}\BibitemShut {NoStop}%
\bibitem [{\citenamefont {Bi}\ \emph {et~al.}(2021)\citenamefont {Bi},
  \citenamefont {Wang}, \citenamefont {Xu}, \citenamefont {Chen}, \citenamefont
  {Lan},\ and\ \citenamefont {Lei}}]{Bi_AdvOptMater_2021}%
  \BibitemOpen
  \bibfield  {author} {\bibinfo {author} {\bibfnamefont {K.}~\bibnamefont
  {Bi}}, \bibinfo {author} {\bibfnamefont {Q.}~\bibnamefont {Wang}}, \bibinfo
  {author} {\bibfnamefont {J.}~\bibnamefont {Xu}}, \bibinfo {author}
  {\bibfnamefont {L.}~\bibnamefont {Chen}}, \bibinfo {author} {\bibfnamefont
  {C.}~\bibnamefont {Lan}},\ and\ \bibinfo {author} {\bibfnamefont
  {M.}~\bibnamefont {Lei}},\ }\bibfield  {title} {\bibinfo {title}
  {All-dielectric metamaterial fabrication techniques},\ }\href
  {https://doi.org/https://doi.org/10.1002/adom.202001474} {\bibfield
  {journal} {\bibinfo  {journal} {Adv. Opt. Mater.}\ }\textbf {\bibinfo
  {volume} {9}},\ \bibinfo {pages} {2001474} (\bibinfo {year}
  {2021})}\BibitemShut {NoStop}%
\bibitem [{\citenamefont {Evlyukhin}\ \emph {et~al.}(2021)\citenamefont
  {Evlyukhin}, \citenamefont {Poleva}, \citenamefont {Prokhorov}, \citenamefont
  {Baryshnikova}, \citenamefont {Miroshnichenko},\ and\ \citenamefont
  {Chichkov}}]{evlyukhin2021polarization}%
  \BibitemOpen
  \bibfield  {author} {\bibinfo {author} {\bibfnamefont {A.~B.}\ \bibnamefont
  {Evlyukhin}}, \bibinfo {author} {\bibfnamefont {M.~A.}\ \bibnamefont
  {Poleva}}, \bibinfo {author} {\bibfnamefont {A.~V.}\ \bibnamefont
  {Prokhorov}}, \bibinfo {author} {\bibfnamefont {K.~V.}\ \bibnamefont
  {Baryshnikova}}, \bibinfo {author} {\bibfnamefont {A.~E.}\ \bibnamefont
  {Miroshnichenko}},\ and\ \bibinfo {author} {\bibfnamefont {B.~N.}\
  \bibnamefont {Chichkov}},\ }\bibfield  {title} {\bibinfo {title}
  {Polarization switching between electric and magnetic quasi-trapped modes in
  bianisotropic all-dielectric metasurfaces},\ }\href
  {https://doi.org/10.1002/lpor.202100206} {\bibfield  {journal} {\bibinfo
  {journal} {Laser Photonics Rev.}\ }\textbf {\bibinfo {volume} {15}},\
  \bibinfo {pages} {2100206} (\bibinfo {year} {2021})}\BibitemShut {NoStop}%
\bibitem [{\citenamefont {Prokhorov}\ \emph
  {et~al.}(2022{\natexlab{a}})\citenamefont {Prokhorov}, \citenamefont
  {Shesterikov}, \citenamefont {Gubin}, \citenamefont {Volkov},\ and\
  \citenamefont {Evlyukhin}}]{Prokhorov_PhysRevB_2022}%
  \BibitemOpen
  \bibfield  {author} {\bibinfo {author} {\bibfnamefont {A.~V.}\ \bibnamefont
  {Prokhorov}}, \bibinfo {author} {\bibfnamefont {A.~V.}\ \bibnamefont
  {Shesterikov}}, \bibinfo {author} {\bibfnamefont {M.~Y.}\ \bibnamefont
  {Gubin}}, \bibinfo {author} {\bibfnamefont {V.~S.}\ \bibnamefont {Volkov}},\
  and\ \bibinfo {author} {\bibfnamefont {A.~B.}\ \bibnamefont {Evlyukhin}},\
  }\bibfield  {title} {\bibinfo {title} {Quasitrapped modes in metasurfaces of
  anisotropic {${\mathrm{MoS}}_{2}$} nanoparticles for absorption and
  polarization control in the telecom wavelength range},\ }\href
  {https://doi.org/10.1103/PhysRevB.106.035412} {\bibfield  {journal} {\bibinfo
   {journal} {Phys. Rev. B}\ }\textbf {\bibinfo {volume} {106}},\ \bibinfo
  {pages} {035412} (\bibinfo {year} {2022}{\natexlab{a}})}\BibitemShut
  {NoStop}%
\bibitem [{\citenamefont {Prokhorov}\ \emph
  {et~al.}(2022{\natexlab{b}})\citenamefont {Prokhorov}, \citenamefont
  {Terekhov}, \citenamefont {Gubin}, \citenamefont {Shesterikov}, \citenamefont
  {Ni}, \citenamefont {Tuz},\ and\ \citenamefont
  {Evlyukhin}}]{Prokhorov_acsphotonics_2022}%
  \BibitemOpen
  \bibfield  {author} {\bibinfo {author} {\bibfnamefont {A.~V.}\ \bibnamefont
  {Prokhorov}}, \bibinfo {author} {\bibfnamefont {P.~D.}\ \bibnamefont
  {Terekhov}}, \bibinfo {author} {\bibfnamefont {M.~Y.}\ \bibnamefont {Gubin}},
  \bibinfo {author} {\bibfnamefont {A.~V.}\ \bibnamefont {Shesterikov}},
  \bibinfo {author} {\bibfnamefont {X.}~\bibnamefont {Ni}}, \bibinfo {author}
  {\bibfnamefont {V.~R.}\ \bibnamefont {Tuz}},\ and\ \bibinfo {author}
  {\bibfnamefont {A.~B.}\ \bibnamefont {Evlyukhin}},\ }\bibfield  {title}
  {\bibinfo {title} {Resonant light trapping via lattice-induced multipole
  coupling in symmetrical metasurfaces},\ }\href
  {https://doi.org/10.1021/acsphotonics.2c01066} {\bibfield  {journal}
  {\bibinfo  {journal} {ACS Photonics}\ }\textbf {\bibinfo {volume} {9}},\
  \bibinfo {pages} {3869} (\bibinfo {year} {2022}{\natexlab{b}})}\BibitemShut
  {NoStop}%
\bibitem [{\citenamefont {Qin}\ \emph {et~al.}(2022)\citenamefont {Qin},
  \citenamefont {Duan}, \citenamefont {Xiao}, \citenamefont {Liu},
  \citenamefont {Yu}, \citenamefont {Wang},\ and\ \citenamefont
  {Liao}}]{qin2022strong}%
  \BibitemOpen
  \bibfield  {author} {\bibinfo {author} {\bibfnamefont {M.}~\bibnamefont
  {Qin}}, \bibinfo {author} {\bibfnamefont {J.}~\bibnamefont {Duan}}, \bibinfo
  {author} {\bibfnamefont {S.}~\bibnamefont {Xiao}}, \bibinfo {author}
  {\bibfnamefont {W.}~\bibnamefont {Liu}}, \bibinfo {author} {\bibfnamefont
  {T.}~\bibnamefont {Yu}}, \bibinfo {author} {\bibfnamefont {T.}~\bibnamefont
  {Wang}},\ and\ \bibinfo {author} {\bibfnamefont {Q.}~\bibnamefont {Liao}},\
  }\bibfield  {title} {\bibinfo {title} {Strong coupling between excitons and
  quasi-bound states in the continuum in the bulk transition metal
  dichalcogenides},\ }\href {https://arxiv.org/pdf/2209.00416} {\bibfield
  {journal} {\bibinfo  {journal} {arXiv preprint arXiv:2209.00416}\ } (\bibinfo
  {year} {2022})}\BibitemShut {NoStop}%
\bibitem [{\citenamefont {Ushkov}\ \emph {et~al.}(2022)\citenamefont {Ushkov},
  \citenamefont {Ermolaev}, \citenamefont {Vyshnevyy}, \citenamefont {Baranov},
  \citenamefont {Arsenin},\ and\ \citenamefont
  {Volkov}}]{Ushkov_PhysRevB_2022}%
  \BibitemOpen
  \bibfield  {author} {\bibinfo {author} {\bibfnamefont {A.~A.}\ \bibnamefont
  {Ushkov}}, \bibinfo {author} {\bibfnamefont {G.~A.}\ \bibnamefont
  {Ermolaev}}, \bibinfo {author} {\bibfnamefont {A.~A.}\ \bibnamefont
  {Vyshnevyy}}, \bibinfo {author} {\bibfnamefont {D.~G.}\ \bibnamefont
  {Baranov}}, \bibinfo {author} {\bibfnamefont {A.~V.}\ \bibnamefont
  {Arsenin}},\ and\ \bibinfo {author} {\bibfnamefont {V.~S.}\ \bibnamefont
  {Volkov}},\ }\bibfield  {title} {\bibinfo {title} {Anapole states and
  scattering deflection effects in anisotropic van der {Waals} nanoparticles},\
  }\href {https://doi.org/10.1103/PhysRevB.106.195302} {\bibfield  {journal}
  {\bibinfo  {journal} {Phys. Rev. B}\ }\textbf {\bibinfo {volume} {106}},\
  \bibinfo {pages} {195302} (\bibinfo {year} {2022})}\BibitemShut {NoStop}%
\bibitem [{\citenamefont {Snitzer}(1961)}]{Snitzer_JOSA_1961}%
  \BibitemOpen
  \bibfield  {author} {\bibinfo {author} {\bibfnamefont {E.}~\bibnamefont
  {Snitzer}},\ }\bibfield  {title} {\bibinfo {title} {Cylindrical dielectric
  waveguide modes},\ }\href {https://doi.org/10.1364/JOSA.51.000491} {\bibfield
   {journal} {\bibinfo  {journal} {J. Opt. Soc. Am.}\ }\textbf {\bibinfo
  {volume} {51}},\ \bibinfo {pages} {491} (\bibinfo {year} {1961})}\BibitemShut
  {NoStop}%
\bibitem [{\citenamefont {Mongia}\ and\ \citenamefont
  {Bhartia}(1994)}]{Mongia_1994}%
  \BibitemOpen
  \bibfield  {author} {\bibinfo {author} {\bibfnamefont {R.~K.}\ \bibnamefont
  {Mongia}}\ and\ \bibinfo {author} {\bibfnamefont {P.}~\bibnamefont
  {Bhartia}},\ }\bibfield  {title} {\bibinfo {title} {Dielectric resonator
  antennas—a review and general design relations for resonant frequency and
  bandwidth},\ }\href {https://doi.org/10.1002/mmce.4570040304} {\bibfield
  {journal} {\bibinfo  {journal} {Int. J. RF Microw. Comput. Aided Eng.}\
  }\textbf {\bibinfo {volume} {4}},\ \bibinfo {pages} {230} (\bibinfo {year}
  {1994})}\BibitemShut {NoStop}%
\bibitem [{\citenamefont {Sayanskiy}\ \emph {et~al.}(2019)\citenamefont
  {Sayanskiy}, \citenamefont {Kupriianov}, \citenamefont {Xu}, \citenamefont
  {Kapitanova}, \citenamefont {Dmitriev}, \citenamefont {Khardikov},\ and\
  \citenamefont {Tuz}}]{Tuz_PhysRevB_2019}%
  \BibitemOpen
  \bibfield  {author} {\bibinfo {author} {\bibfnamefont {A.}~\bibnamefont
  {Sayanskiy}}, \bibinfo {author} {\bibfnamefont {A.~S.}\ \bibnamefont
  {Kupriianov}}, \bibinfo {author} {\bibfnamefont {S.}~\bibnamefont {Xu}},
  \bibinfo {author} {\bibfnamefont {P.}~\bibnamefont {Kapitanova}}, \bibinfo
  {author} {\bibfnamefont {V.}~\bibnamefont {Dmitriev}}, \bibinfo {author}
  {\bibfnamefont {V.~V.}\ \bibnamefont {Khardikov}},\ and\ \bibinfo {author}
  {\bibfnamefont {V.~R.}\ \bibnamefont {Tuz}},\ }\bibfield  {title} {\bibinfo
  {title} {Controlling high-{$Q$} trapped modes in polarization-insensitive
  all-dielectric metasurfaces},\ }\href
  {https://doi.org/10.1103/PhysRevB.99.085306} {\bibfield  {journal} {\bibinfo
  {journal} {Phys. Rev. B}\ }\textbf {\bibinfo {volume} {99}},\ \bibinfo
  {pages} {085306} (\bibinfo {year} {2019})}\BibitemShut {NoStop}%
\bibitem [{\citenamefont {Vaity}\ \emph
  {et~al.}(2022{\natexlab{a}})\citenamefont {Vaity}, \citenamefont {Gupta},
  \citenamefont {Kala}, \citenamefont {Dutta~Gupta}, \citenamefont {Kivshar},
  \citenamefont {Tuz},\ and\ \citenamefont {Achanta}}]{Pravin_AdvPhRes_2022}%
  \BibitemOpen
  \bibfield  {author} {\bibinfo {author} {\bibfnamefont {P.}~\bibnamefont
  {Vaity}}, \bibinfo {author} {\bibfnamefont {H.}~\bibnamefont {Gupta}},
  \bibinfo {author} {\bibfnamefont {A.}~\bibnamefont {Kala}}, \bibinfo {author}
  {\bibfnamefont {S.}~\bibnamefont {Dutta~Gupta}}, \bibinfo {author}
  {\bibfnamefont {Y.~S.}\ \bibnamefont {Kivshar}}, \bibinfo {author}
  {\bibfnamefont {V.~R.}\ \bibnamefont {Tuz}},\ and\ \bibinfo {author}
  {\bibfnamefont {V.~G.}\ \bibnamefont {Achanta}},\ }\bibfield  {title}
  {\bibinfo {title} {Polarization-independent quasibound states in the
  continuum},\ }\href {https://doi.org/https://doi.org/10.1002/adpr.202100144}
  {\bibfield  {journal} {\bibinfo  {journal} {Adv. Photonics Res.}\ }\textbf
  {\bibinfo {volume} {3}},\ \bibinfo {pages} {2100144} (\bibinfo {year}
  {2022}{\natexlab{a}})}\BibitemShut {NoStop}%
\bibitem [{\citenamefont {Tuz}\ \emph {et~al.}(2020)\citenamefont {Tuz},
  \citenamefont {Yu}, \citenamefont {Dmitriev},\ and\ \citenamefont
  {Kivshar}}]{Tuz_PhysRevApplied_2020}%
  \BibitemOpen
  \bibfield  {author} {\bibinfo {author} {\bibfnamefont {V.~R.}\ \bibnamefont
  {Tuz}}, \bibinfo {author} {\bibfnamefont {P.}~\bibnamefont {Yu}}, \bibinfo
  {author} {\bibfnamefont {V.}~\bibnamefont {Dmitriev}},\ and\ \bibinfo
  {author} {\bibfnamefont {Y.~S.}\ \bibnamefont {Kivshar}},\ }\bibfield
  {title} {\bibinfo {title} {Magnetic dipole ordering in resonant dielectric
  metasurfaces},\ }\href {https://doi.org/10.1103/PhysRevApplied.13.044003}
  {\bibfield  {journal} {\bibinfo  {journal} {Phys. Rev. Applied}\ }\textbf
  {\bibinfo {volume} {13}},\ \bibinfo {pages} {044003} (\bibinfo {year}
  {2020})}\BibitemShut {NoStop}%
\bibitem [{\citenamefont {Tuz}\ and\ \citenamefont
  {Evlyukhin}(2021)}]{Tuz_Nanophot_2021}%
  \BibitemOpen
  \bibfield  {author} {\bibinfo {author} {\bibfnamefont {V.~R.}\ \bibnamefont
  {Tuz}}\ and\ \bibinfo {author} {\bibfnamefont {A.~B.}\ \bibnamefont
  {Evlyukhin}},\ }\bibfield  {title} {\bibinfo {title}
  {Polarization-independent anapole response of a trimer-based dielectric
  metasurface},\ }\href {https://doi.org/10.1515/nanoph-2021-0315} {\bibfield
  {journal} {\bibinfo  {journal} {Nanophotonics}\ }\textbf {\bibinfo {volume}
  {10}},\ \bibinfo {pages} {4373} (\bibinfo {year} {2021})}\BibitemShut
  {NoStop}%
\bibitem [{\citenamefont {Vaity}\ \emph
  {et~al.}(2022{\natexlab{b}})\citenamefont {Vaity}, \citenamefont {Gupta},
  \citenamefont {Kala}, \citenamefont {Dutta~Gupta}, \citenamefont {Kivshar},
  \citenamefont {Tuz},\ and\ \citenamefont {Gopal~Achanta}}]{Tuz_Advphot_2022}%
  \BibitemOpen
  \bibfield  {author} {\bibinfo {author} {\bibfnamefont {P.}~\bibnamefont
  {Vaity}}, \bibinfo {author} {\bibfnamefont {H.}~\bibnamefont {Gupta}},
  \bibinfo {author} {\bibfnamefont {A.}~\bibnamefont {Kala}}, \bibinfo {author}
  {\bibfnamefont {S.}~\bibnamefont {Dutta~Gupta}}, \bibinfo {author}
  {\bibfnamefont {Y.~S.}\ \bibnamefont {Kivshar}}, \bibinfo {author}
  {\bibfnamefont {V.~R.}\ \bibnamefont {Tuz}},\ and\ \bibinfo {author}
  {\bibfnamefont {V.}~\bibnamefont {Gopal~Achanta}},\ }\bibfield  {title}
  {\bibinfo {title} {Polarization-independent quasi-bound states in the
  continuum},\ }\href {https://doi.org/10.1002/adpr.202100144} {\bibfield
  {journal} {\bibinfo  {journal} {Adv. Photonics Res.}\ }\textbf {\bibinfo
  {volume} {3}},\ \bibinfo {pages} {2100144} (\bibinfo {year}
  {2022}{\natexlab{b}})}\BibitemShut {NoStop}%
\bibitem [{\citenamefont {Tonning}(1982)}]{Tonning_MTT_1982}%
  \BibitemOpen
  \bibfield  {author} {\bibinfo {author} {\bibfnamefont {A.}~\bibnamefont
  {Tonning}},\ }\bibfield  {title} {\bibinfo {title} {Circularly symmetric
  optical waveguide with strong anisotropy},\ }\href
  {https://doi.org/10.1109/TMTT.1982.1131138} {\bibfield  {journal} {\bibinfo
  {journal} {IEEE Trans. Microwave Theory Techn.}\ }\textbf {\bibinfo {volume}
  {30}},\ \bibinfo {pages} {790} (\bibinfo {year} {1982})}\BibitemShut
  {NoStop}%
\bibitem [{\citenamefont {Paul}\ and\ \citenamefont
  {Shevgaonkar}(1981)}]{Paul_RS_1981}%
  \BibitemOpen
  \bibfield  {author} {\bibinfo {author} {\bibfnamefont {D.~K.}\ \bibnamefont
  {Paul}}\ and\ \bibinfo {author} {\bibfnamefont {R.~K.}\ \bibnamefont
  {Shevgaonkar}},\ }\bibfield  {title} {\bibinfo {title} {Multimode propagation
  in anisotropic optical waveguides},\ }\href
  {https://doi.org/10.1029/RS016i004p00525} {\bibfield  {journal} {\bibinfo
  {journal} {Radio Sci.}\ }\textbf {\bibinfo {volume} {16}},\ \bibinfo {pages}
  {525} (\bibinfo {year} {1981})}\BibitemShut {NoStop}%
\bibitem [{\citenamefont {Yu}\ \emph {et~al.}(2018)\citenamefont {Yu},
  \citenamefont {Fesenko},\ and\ \citenamefont {Tuz}}]{Yu_Nanophot_2018}%
  \BibitemOpen
  \bibfield  {author} {\bibinfo {author} {\bibfnamefont {P.}~\bibnamefont
  {Yu}}, \bibinfo {author} {\bibfnamefont {V.~I.}\ \bibnamefont {Fesenko}},\
  and\ \bibinfo {author} {\bibfnamefont {V.~R.}\ \bibnamefont {Tuz}},\
  }\bibfield  {title} {\bibinfo {title} {Dispersion features of complex waves
  in a graphene-coated semiconductor nanowire},\ }\href
  {https://doi.org/doi:10.1515/nanoph-2018-0026} {\bibfield  {journal}
  {\bibinfo  {journal} {Nanophotonics}\ }\textbf {\bibinfo {volume} {7}},\
  \bibinfo {pages} {925} (\bibinfo {year} {2018})}\BibitemShut {NoStop}%
\bibitem [{\citenamefont {Vandenbulcke}\ and\ \citenamefont
  {Lagasse}(1976)}]{Vandenbulcke_EL_1976}%
  \BibitemOpen
  \bibfield  {author} {\bibinfo {author} {\bibfnamefont {P.}~\bibnamefont
  {Vandenbulcke}}\ and\ \bibinfo {author} {\bibfnamefont {P.~E.}\ \bibnamefont
  {Lagasse}},\ }\bibfield  {title} {\bibinfo {title} {Eigenmode analysis of
  anisotropic optical fibers or integrated optical waveguides},\ }\href
  {https://doi.org/10.1049/el:19760095} {\bibfield  {journal} {\bibinfo
  {journal} {Electron. Lett.}\ }\textbf {\bibinfo {volume} {12}},\ \bibinfo
  {pages} {120} (\bibinfo {year} {1976})}\BibitemShut {NoStop}%
\bibitem [{\citenamefont {Koshiba}\ \emph {et~al.}(1986)\citenamefont
  {Koshiba}, \citenamefont {Hayata},\ and\ \citenamefont
  {Suzuki}}]{Koshiba_JLT_1986}%
  \BibitemOpen
  \bibfield  {author} {\bibinfo {author} {\bibfnamefont {M.}~\bibnamefont
  {Koshiba}}, \bibinfo {author} {\bibfnamefont {K.}~\bibnamefont {Hayata}},\
  and\ \bibinfo {author} {\bibfnamefont {M.}~\bibnamefont {Suzuki}},\
  }\bibfield  {title} {\bibinfo {title} {Finite-element solution of anisotropic
  waveguides with arbitrary tensor permittivity},\ }\href
  {https://doi.org/10.1109/JLT.1986.1074687} {\bibfield  {journal} {\bibinfo
  {journal} {J. Lightwave Technol.}\ }\textbf {\bibinfo {volume} {4}},\
  \bibinfo {pages} {121} (\bibinfo {year} {1986})}\BibitemShut {NoStop}%
\bibitem [{\citenamefont {Bagatskaya}\ \emph {et~al.}(2005)\citenamefont
  {Bagatskaya}, \citenamefont {Fesenko},\ and\ \citenamefont
  {Shulga}}]{Bagatskaya_TRE_2005}%
  \BibitemOpen
  \bibfield  {author} {\bibinfo {author} {\bibfnamefont {O.~V.}\ \bibnamefont
  {Bagatskaya}}, \bibinfo {author} {\bibfnamefont {V.~I.}\ \bibnamefont
  {Fesenko}},\ and\ \bibinfo {author} {\bibfnamefont {S.~N.}\ \bibnamefont
  {Shulga}},\ }\bibfield  {title} {\bibinfo {title} {Electric field lines in a
  rectangular waveguide with an inhomogeneous anisotropic insert},\ }\href
  {https://doi.org/10.1615/TelecomRadEng.v63.i1.10} {\bibfield  {journal}
  {\bibinfo  {journal} {Telecommun. Radio Eng.}\ }\textbf {\bibinfo {volume}
  {63}},\ \bibinfo {pages} {1} (\bibinfo {year} {2005})}\BibitemShut {NoStop}%
\bibitem [{\citenamefont {Dai}\ and\ \citenamefont
  {Jen}(1991)}]{Dai_JOSA_1991}%
  \BibitemOpen
  \bibfield  {author} {\bibinfo {author} {\bibfnamefont {J.~D.}\ \bibnamefont
  {Dai}}\ and\ \bibinfo {author} {\bibfnamefont {C.~K.}\ \bibnamefont {Jen}},\
  }\bibfield  {title} {\bibinfo {title} {Analysis of cladded uniaxial
  single-crystal fibers},\ }\href {https://doi.org/10.1364/JOSAA.8.002021}
  {\bibfield  {journal} {\bibinfo  {journal} {J. Opt. Soc. Am. A}\ }\textbf
  {\bibinfo {volume} {8}},\ \bibinfo {pages} {2021} (\bibinfo {year}
  {1991})}\BibitemShut {NoStop}%
\bibitem [{\citenamefont {Dai}\ \emph {et~al.}(1992)\citenamefont {Dai},
  \citenamefont {de~Oliveira},\ and\ \citenamefont {Jen}}]{Dai_JOSA_1992}%
  \BibitemOpen
  \bibfield  {author} {\bibinfo {author} {\bibfnamefont {J.~D.}\ \bibnamefont
  {Dai}}, \bibinfo {author} {\bibfnamefont {C.~A.~S.}\ \bibnamefont
  {de~Oliveira}},\ and\ \bibinfo {author} {\bibfnamefont {C.~K.}\ \bibnamefont
  {Jen}},\ }\bibfield  {title} {\bibinfo {title} {Dispersion characteristics of
  cladded biaxial optical fibers},\ }\href
  {https://doi.org/10.1364/JOSAA.9.001564} {\bibfield  {journal} {\bibinfo
  {journal} {J. Opt. Soc. Am. A}\ }\textbf {\bibinfo {volume} {9}},\ \bibinfo
  {pages} {1564} (\bibinfo {year} {1992})}\BibitemShut {NoStop}%
\bibitem [{\citenamefont {Evlyukhin}\ and\ \citenamefont
  {Chichkov}(2019)}]{evlyukhin2019multipole}%
  \BibitemOpen
  \bibfield  {author} {\bibinfo {author} {\bibfnamefont {A.~B.}\ \bibnamefont
  {Evlyukhin}}\ and\ \bibinfo {author} {\bibfnamefont {B.~N.}\ \bibnamefont
  {Chichkov}},\ }\bibfield  {title} {\bibinfo {title} {Multipole decompositions
  for directional light scattering},\ }\href
  {https://doi.org/10.1103/PhysRevB.100.125415} {\bibfield  {journal} {\bibinfo
   {journal} {Phys. Rev. B}\ }\textbf {\bibinfo {volume} {100}},\ \bibinfo
  {pages} {125415} (\bibinfo {year} {2019})}\BibitemShut {NoStop}%
\end{thebibliography}%

\end{document}